\documentclass[aps,prb,twocolumn,floatfix,superscriptaddress]{revtex4-2}

\usepackage{amssymb}
\usepackage{amsmath}
\usepackage{graphicx}

\begin{document}

\title{Model for charge carrier spectra in topological semimetals of the TaAs family}

\author{G.\ P.\ Mikitik}

\affiliation{B.~Verkin Institute for Low Temperature Physics \&
Engineering of the National Academy of Sciences of Ukraine, Kharkiv 61103, Ukraine}

\author{Yu.\ V.\ Sharlai}

\affiliation{B.~Verkin Institute for Low Temperature Physics \&
Engineering of the National Academy of Sciences of Ukraine, Kharkiv 61103, Ukraine}

\affiliation{Institute of Low Temperature and Structure Research, Polish Academy of  Sciences, 50-422 Wroc{\l}aw, Poland}

\begin{abstract}
We propose a four-band model describing the electron energy spectra near the Weyl points in the topological semimetals of the TaAs family (TaAs, TaP, NbAs, NbP). This model takes into account the fact that
these Weyl points result from the band-contact lines which would exist in the mirror-reflection planes of these materials if
the spin-orbit interaction were absent in them. Within this model, we obtain conditions for the existence of the Weyl points, determine their positions in the Brillouin zone, and derive the explicit formula for dispersion of the bands along the straight line connecting the two close Weyl points with opposite topological charges.   Using NbP as an example, the values of the parameters defining the  model spectrum are found. The obtained results show that for the semimetals of the TaAs family, the charge-carriers spectrum in the vicinity of the two close Weyl points can be analyzed without complex band-structure calculations.
\end{abstract}

\maketitle

\section{Introduction}

In Weyl semimetals, two singly degenerate bands of charge carriers contact at discrete (Weyl) points of the Brillouin zone and disperse linearly in all directions around these points. These spin-nondegenerate bands occur in crystals with a significant spin-orbit interaction if either spatial inversion or time reversal symmetry is broken in them. Here we will consider only the noncentrosymmetric Weyl semimetals. This class of the topological semimetals include  the TaAs family (TaAs, TaP, NbAs, and NbP) that was intensively investigated in recent years; see, e.g., review articles \cite{armit,lv} and references therein.

To describe the dispersion of the bands near a Weyl point, the simple model spectrum is frequently used,
\[
 \varepsilon=\varepsilon^W+{\bf a}{\bf p}\pm \sqrt{v_x^2p_x^2 + v_y^2p_y^2 +v_z^2p_z^2},
\]
where $p_i$ ($i=x,y,z$) are the components of the quasimomentum, the velocities $v_i$ specify the splitting of the bands at the Weyl point ${\bf p}=0$ that has the energy $\varepsilon^W$, and the constant vector ${\bf a}=(a_x,a_y,a_z)$ with the dimension of the velocity determines the so-called tilt of the spectrum. Moreover, the use of the additional simplification ${\bf a}=0$ is not rare in publications. However, only pairs of Weyl points with the opposite topological charges (the Chern numbers) can arise in the Brillouin zone of crystals, and the distance between these points of each pair is relatively small. This means that the region of the Brillouin zone where the linear dispersion of the bands really occurs is small, and a deviation from the linear dispersion can manifest itself even for a small difference between the Fermi energy $\varepsilon_F$ and the energy $\varepsilon^W$ of the Weyl points in the pair. Therefore, in order to correctly find the Fermi surface and its extremal cross sections, calculate the Landau levels and various physical quantities for the Weyl semimetals in a magnetic field, it may be necessary to consider the energy spectrum of the pair of the Weyl points \cite{jiang} (see also recent papers \cite{polatkan,zhao,moon,balduini}). Moreover, if  $|\varepsilon_F- \varepsilon^W|$ is larger than the energy barrier separating the points, for the correct description of the charge-carriers spectrum, it is necessary to take into account four close energy bands that exist near the Weyl points. These four bands result from the nodal line (lying in the mirror-reflection plane) which would exist in these semimetals if the spin-orbit were absent in them \cite{h.weng}. The crossings of two bands from this four just produce the pair of the Weyl points. Due to this complexity of the charge-carriers spectrum, numerical calculations of the total electron-band structure of Weyl semimetals are commonly used to find the Fermi surface and its characteristics in these materials \cite{lee,klotz,wu,grassano18,grassano}.

However,  Weng {\it et al}.\ \cite{h.weng} indicated the general form of the Hamiltonian that describes the four bands originating from the band-contact line lying in the mirror-reflection plane (the plane is also assumed to contain the twofold or fourfold symmetry axis). Such planes just exist in semimetals of the TaAs family. With this general form, we formulate the four-band model describing the spectrum of the charge carriers near the two close Weyl points in these materials (Sec.~\ref{spectrum}). Within this model, we find conditions for the existence of the Weyl points and explicit analytical formulas for $\varepsilon^W$ and for the dispersion of the four bands along the line connecting the close Weyl points in the Brillouin zone. It also follows from the results of this section that the four bands and  the cross-sectional areas of the Fermi surface can be found by numerically solving a quartic equation, avoiding the complicated band-structure calculations. In other words, the model makes it possible to simply analyze various experimental data obtained with oscillation effects. In Sec.~\ref{NbP}, using NbP as an example, we demonstrate how the parameters of the model can be found  for the so-called \cite{armit} W1 and W2 Weyl points, comparing the characteristic features of the model spectrum with the data of the band-structure calculations for this semimetal \cite{lee}. In Sec.~\ref{discussion}, the obtained results are discussed, and conclusions are presented in Sec.~\ref{conclusions}. Appendixes contain some mathematical details.

\section{Charge-carriers spectra near Weyl points} \label{spectrum}

To describe the  charge-carriers spectra near the Weyl points in the TaAs family of topological semimetals, we  modify the Hamiltonian of Weng {\it et al}.\ \cite{h.weng}. This  Hamiltonian demonstrates how these points result from the band-contact line lying in the mirror-reflection plane, $p_y=0$, when the spin-orbit interaction is ``turned on''. We will consider the following Hamiltonian:
 \begin{eqnarray}\label{1}
\hat H=\left (\begin{array}{cccc} \bar\varepsilon_0 +d_z & M_+ & T_1 & U \\ M_+^*& \bar\varepsilon_0 +d_z &U & T_2 \\ T_1^* & U^* & \bar\varepsilon_0 -d_z & M_- \\ U^*& T_2^* & M_-^* & \bar\varepsilon_0 -d_z \\
\end{array} \right),
 \end{eqnarray}
where $\bar\varepsilon_0=\bar\varepsilon_0(p_x,p_z)$, $d_z=d_z(p_x,p_z)$ are some functions  of the quasimomentum in the plane $p_y=0$, $p_i$ are the components of the quasimomentum,
 \begin{eqnarray}\label{2}
T_1&=&(t_1-it_2)p_y +(m_4-im_5), \nonumber \\
T_2&=&(t_1-it_2)p_y -(m_4-im_5), \\
U&=&m_3-im_6, \nonumber \\
M_{\pm}&=&i(m_1\pm m_2), \nonumber
  \end{eqnarray}
$m_1$, $m_2$, $m_3$, $m_4$, $m_5$, $m_6$ are relatively small parameters proportional to the strength of the spin-orbit interaction, and $t_1=t_1(p_x,p_z)$, $t_2=t_2(p_x,p_z)$ are the matrix elements of the velocity operator. The functions $\bar\varepsilon_0(p_x,p_z)$, $d_z(p_x,p_z)$, $t_1(p_x,p_z)$, $t_2(p_x,p_z)$ exist when the spin-orbit coupling is absent, and so they are practically independent of this coupling if it is relatively weak. As compared to Ref.~\cite{h.weng}, we have added the term $\bar\varepsilon_0$ to $\hat H$.

\begin{figure}[t] % %%%%%%%%%%%%%%%%%%%%%%%%%%%%%%%%%%%%%
 \centering  \vspace{+9 pt}
\includegraphics[scale=0.99]{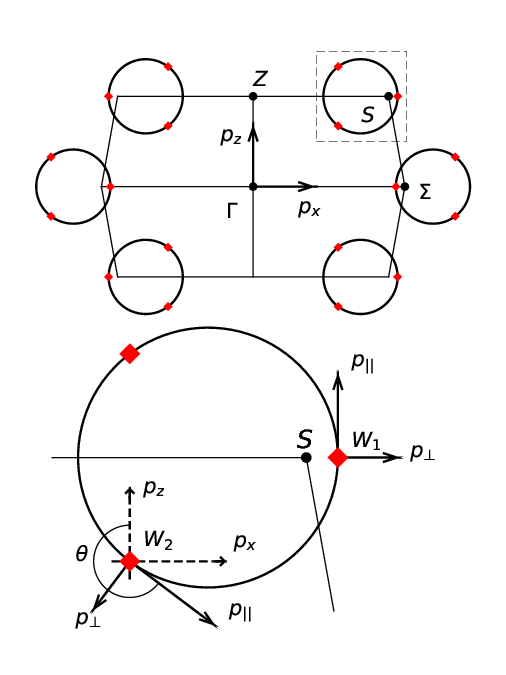}
\caption{\label{fig1} Cross section (hexagon) of the first Brillouin zone of NbP by the mirror-reflection plane $p_y=0$, and the band-contact rings in this plane. The red rhombi depict the projections of the Weyl points onto this plane. The upper right  ring (enclosed by the dashed square) is also shown in an enlarged scale. For this ring, the W1 and W2 Weyl points are marked, near which the energy bands were calculated in Ref.~\cite{lee} (see subsequent Figs.~\ref{fig2}, \ref{fig3}, \ref{fig5}-\ref{fig8}, and Tables \ref{tab1}, \ref{tab3}). For these Weyl points, the angle $\theta$ between the $p_z$ axis and the $p_{\|}$ direction is indicated.
}
\end{figure}

In neglect of the spin-orbit interaction, all $m_i=0$ ($i=1-6$), and hence, $T_1=T_2$, $U=0$, $M_+=M_-=0$. In this case, the Hamiltonian (\ref{1}), (\ref{2}) gives the following dispersion of the two bands (doubly degenerated in spin),
 \[
  \varepsilon({\bf p})=\bar\varepsilon_0\pm \sqrt{d_z^2+p_y^2(t_1^2+t_2^2)}
 \]
A contact line of these two bands lies in the plane $p_y=0$ and is determined by the condition,
 \begin{eqnarray}\label{3}
d_z(p_x,p_z)=0.
\end{eqnarray}
In the semimetals of the TaAs family, this line is a ring \cite{h.weng,lee,klotz,grassano18,grassano} (Fig.~\ref{fig1}). We will consider the electron spectrum in the vicinity of this band-contact line. Then, in the mirror-reflection plane $p_y=0$, it is convenient to introduce the curvilinear coordinate $p_{\parallel}$ along the ring and the local coordinate $p_{\perp}$ perpendicular to the ring at a given point (i.e., $p_{\perp}=0$ at any point of the ring). Now let us further refine  Hamiltonian (\ref{1}):  in the vicinity of the band-contact line, we will use the expansions,
  \begin{eqnarray}\label{4}
d_z&=&a'(p_{\parallel})p_{\perp}, \\
\bar\varepsilon_0&=&\varepsilon_0(p_{\parallel}) +a(p_{\parallel})p_{\perp}. \nonumber
  \end{eqnarray}
The quantity $\varepsilon_0(p_{\parallel})$ describes the energy of the two crossing bands along the line, whereas the term $a(p_{\parallel})p_{\perp}$ leads to the so-called tilt of the Dirac spectrum in the planes perpendicular to the line. The functions $\varepsilon_0(p_{\parallel})$, $a(p_{\parallel})$, $a'(p_{\parallel})$, $t_i(p_{\parallel})$  can  noticeably change along the line. In fact, the above assumptions about the dependences of $\bar\varepsilon_0$ and $d_z$ on $p_{\perp}$ mean that in any plane perpendicular to the line, we use the well-known ${\bf k\cdot p}$ approximation to describe the spectrum of the charge carriers near this line. Note that  one can always vanish $t_1$, choosing the appropriate phases of the wave functions, and so without the loss in generality, we set $t_1=0$ below.

   %%%%%%%%%%%%%%%%%%%%%%%%%%%%%%%%%%%%%%%%%%

When $m_i\neq 0$ ($i=1-6$), the contact of the bands in the plane disappears, and couples of the Weyl points can occur near it.
Diagonalization of the  Hamiltonian (\ref{1}) leads to a fourth-order equation that determines  the   dispersion  of  the four spin-nondegenerate bands of charge carriers. (We will numerate the bands in increasing order of their energies.) Generally speaking, the energies of the bands as functions of the quasimomentum cannot be calculated analytically with such an equation. However, this equation is biquadratic with respect to $p_y$, and  we find the explicit formula for $p_y$ as a function of the energy $\varepsilon$ and $p_x$, $p_z$,
 \begin{eqnarray}\label{5}
t_2^2p_y^2&\!=&\epsilon^2-d_z^2+m_1^2 -m_2^2-m_3^2-m_4^2 \nonumber \\
&+&m_5^2+m_6^2 \pm \sqrt{Y},~~~
\end{eqnarray}
where $\epsilon\equiv \varepsilon-\bar\varepsilon_0$,
\begin{eqnarray}\label{6}
Y\!&=&\!4(m_1^2\!+\!m_5^2\!+\!m_6^2)\!\!\left[\epsilon+\frac{m_2(d_zm_1 +m_3m_5- m_4m_6)}{m_1^2+m_5^2+m_6^2}\right]^2\nonumber \\ &+&Y_{min}, \end{eqnarray}
\begin{eqnarray}\label{7}
Y_{min}\!&=&\!-4(m_1^2\!+\!m_5^2\!+\!m_6^2\!-\!m_2^2) \Big[\frac{(m_3m_6\!+\!m_4m_5)^2}{m_5^2\!+\!m_6^2} \nonumber\\ &+&\!\frac{(m_5^2\!+\!m_6^2)}{(m_1^2\!+\!m_5^2\!+\!m_6^2)}\Big(\!d_z\! +\!\frac{m_1(m_4m_6\!-\!m_3m_5)}{m_5^2+m_6^2}\Big)^2\Big].~~`
\end{eqnarray}
Formulas (\ref{5})--(\ref{7}) give the convenient representation of the dispersion relation, and they can be used to calculate and analyze the electron spectra near the W1 and W2 Weyl points in the TaAs family of the topological semimetals. Note that the parameters $m_i$ change along the line (\ref{3}), but we neglect their dependences on $p_{\perp}$  since the dispersion of the bands in this direction is mainly determined by $a'(p_{\|})$  which is not associated with the spin-orbit interaction and is relatively large. In other words, we take $m_i=m_i(p_{\|})$.

The dependences of $\bar\varepsilon_0$, $d_z$, $t_2$, $m_i$ on the quasimomentum $p_{\|}$ in the mirror-reflection plane must also  satisfy the constraints that follow from the body-centered tetragonal crystalline symmetry of the TaAs family. Specifically, the charge-carriers spectra in the planes $p_z=0$ and $p_z=\pm 2\pi\hbar/c$ ($c$ is parameter of the crystal lattice) are invariant under the two consecutive  transformations: the two-fold rotation about the $p_z$ axis and the time reversal. This symmetry imposes the following restrictions on the above-mentioned dependences in the mirror-reflection plane \cite{h.weng}:
 \begin{eqnarray}\label{8}
  \bar\varepsilon_0(p_x,p_z)&=&\bar\varepsilon_0(p_x,-p_z),\ \  d_z(p_x,p_z)\!=d_z(p_x,-p_z), \nonumber \\
  m_6(p_x,p_z)&=&m_6(p_x,-p_z),\ \ m_4(p_x,p_z)\!=m_4(p_x,-p_z), \nonumber \\
   t_2(p_x,p_z)&=&t_2(p_x,-p_z), \\
   m_3(p_x,p_z)\!&=&\!-m_3(p_x,-p_z),\ m_5(p_x,p_z)\!=\!-m_5(p_x,-p_z), \nonumber \\
   m_1(p_x,p_z)&=&m_1(p_x,-p_z),\ \ m_2(p_x,p_z)\!=m_2(p_x,-p_z). \nonumber
    \end{eqnarray}
Similar relationships can be written for reflections $p_z$ relative to the faces of the Brillouin zone $p_z=\pm 2\pi\hbar/c$.

At the Weyl points in the TaAs family of the topological semimetals, a gap in the spectrum of the two crossing bands is absent. This means that in Eq.~(\ref{5}), the quantity $Y$ has to be nonnegative for all $\varepsilon$, i.e., its minimal value  over $\varepsilon$, $Y_{min}$, cannot be less than zero. When $Y_{min}$ reaches zero, we  have a crossing of the two bands, and a couple of the Weyl points appears. The minimal value of $Y$ reaches zero when the following two conditions are simultaneously fulfilled \cite{com}:
\begin{eqnarray}\label{9}
d_z(p_x,p_z)=a'(p_{\parallel})p_{\perp}=-m_1(p_{\parallel}) \frac{m_4(p_{\parallel})}{m_6(p_{\parallel})}, \\
m_3(p_{\parallel})m_6(p_{\parallel})+ m_4(p_{\parallel})m_5(p_{\parallel})=0. \label{10}
\end{eqnarray}
Since $m_i$ depend on $p_{\parallel}$, Eq.~(\ref{10})  determines the coordinate $p_{\parallel}^W$ in the band-contact line where the couple of the Weyl points can appear, whereas formula (\ref{9}) defines a small shift of these points, $p_{\perp}^W= -m_1m_4/(m_6a')|_{p_{\|}=p_{\|}^W}$, relative to the line. These $p_{\parallel}^W$ and $p_{\perp}^W$ determine the coordinates $p_x^W$ and $p_z^W$ of the couple of the Weyl points. Under conditions (\ref{9}) and (\ref{10}), the $Y$ can be rewritten as follows:
\begin{eqnarray}\label{11}
Y=4(m_1^2+m_5^2+m_6^2)\left[\varepsilon -\bar\varepsilon_0- \frac{m_2m_4}{m_6}\right]^2\!\!\!.~~
\end{eqnarray}
Therefore, setting $Y=0$, we finds the energy $\varepsilon^{W}$ of the two Weyl points,
\begin{eqnarray}\label{12}
\varepsilon^{W}-\bar\varepsilon_0^W
=\frac{m_2m_4}{m_6}|_{p_{\|}=p_{\|}^W},
\end{eqnarray}
where
  \begin{eqnarray}\label{13}
  \bar\varepsilon_0^W=\bar\varepsilon_0(p_x^W,p_z^W)= \varepsilon_0(p_{\|}^W)- \frac{m_1m_4 a}{m_6a'}|_{p_{\|}=p_{\|}^W}.
  \end{eqnarray}
Inserting formulas (\ref{9}), (\ref{12}), and $Y=0$ into Eqs.~(\ref{5}), we arrive at the expression,
 \begin{eqnarray}\label{14}
(t_2p_y^W)^2=(1-\frac{m_4^2}{m_6^2})(m_6^2+m_5^2+m_1^2-m_2^2),
\end{eqnarray}
which determines the coordinate $p_y=\pm p_y^{W}$ of the two Weyl points. Here $t_2=t_2(p_{\|}^W)$ and $m_i=m_i(p_{\|}^W)$. It follows from Eq.~(\ref{14}) that the necessary conditions for the existence of the Weyl points are
 \begin{eqnarray}\label{15}
m_4^2<m_6^2,\ \  {\rm and}\ \  m_2^2< m_6^2+m_5^2+m_1^2,
\end{eqnarray}
(or the opposite inequalities). As will be shown in the next section, one more condition is always fulfilled for the Weyl points in the semimetals of the TaAs family \cite{com1}:
\begin{equation}\label{16}
\frac{m_4}{m_6}\sqrt{m_1^2+m_5^2+m_6^2}<|m_2|.
\end{equation}

 \begin{figure}[t] % %%%%%%%%%%%%%%%%%%%%%%%%%%%%%%%%%%%%%
 \centering  \vspace{+9 pt}
\includegraphics[scale=0.44]{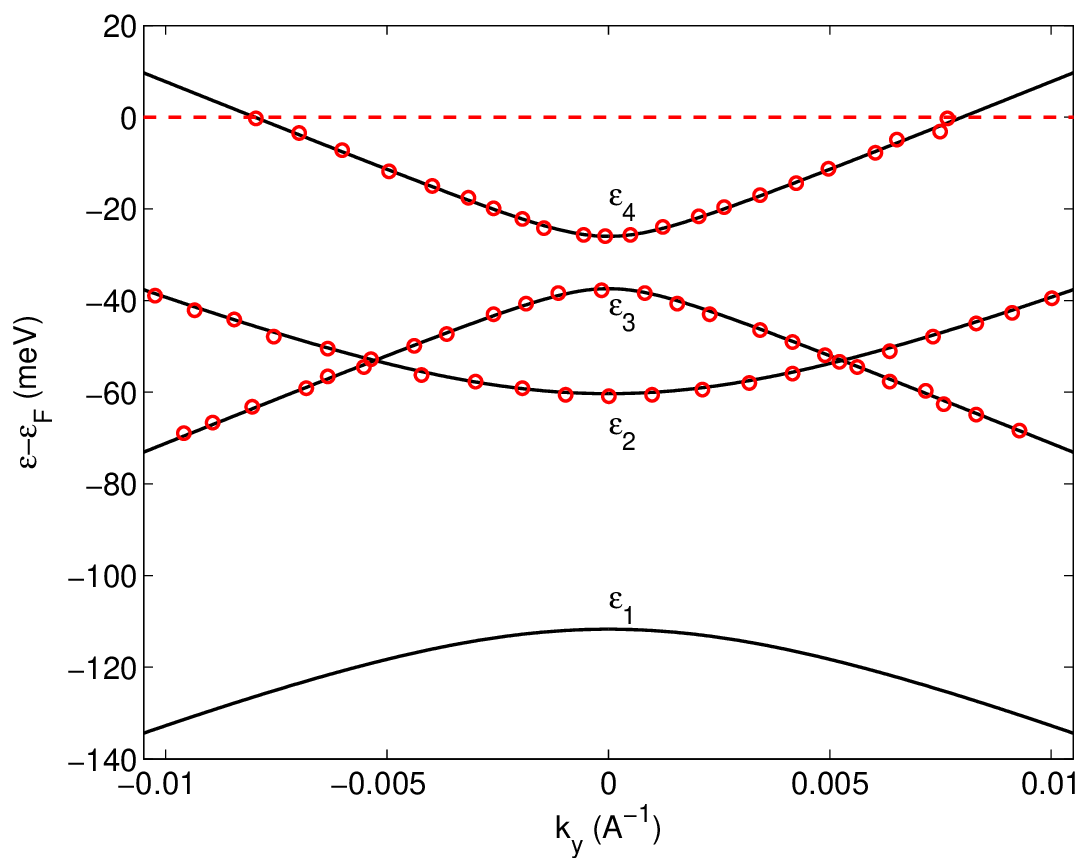}
\caption{\label{fig2} Dispersion of the four bands along the $p_y$ direction ($k_y=p_y/\hbar$) at $p_x=p_x^W$, $p_z=p_z^W$, Eq.~(\ref{17}). The point ($p_x^W$,$p_z^W$) in the mirror-reflection plane $p_y=0$ is the projection  of the two Weyl points onto this  plane. The bands are plotted for $m_3=m_5=0$ and the values of the other parameters presented in Table \ref{tab2}. (Interestingly, the same bands are obtained if the values of $m_2$ and $m_4/m_6$ are replaced by $m_2=10$ meV and $m_4/m_6=0.578$.) The red cycles are the data of Fig.~8a in Ref.~\cite{lee} for the W1 point in NbP (see Fig.~\ref{fig1}) ; $\varepsilon_i$ mark the energy bands. All the energies are measured from the Fermi level (red dashed line).
}
\end{figure}   %%%%%%%%%%%%%%%%%%%%%%%%%%%%%%%%%%%%%%%%%%

At $p_z=0$ or $p_z=\pm 2\pi\hbar/c$, equations (\ref{8}) give $m_3=m_5=0$, and condition (\ref{10}) is fulfilled due to the symmetry. In other words, in the semimetals of TaAs family, the Weyl points can exist when band-contact rings (\ref{3}) lying in the $p_x-p_z$ plane cross the lines $p_z=0, \pm 2\pi\hbar/c$ (Fig.~\ref{fig1}). Such Weyl points are called the W1 points.  Interestingly, each of the band-contact rings crosses one of these straight lines  twice. The absence of the Weyl points for the second crossing means that one of the conditions (\ref{15}) fails at the point of this  crossing. The so-called W2 points lie at $p_z\neq 0, \pm 2\pi\hbar/c$, and their positions on the band-contact ring is determined by Eq.~(\ref{10}).

Let us presents several results that follows from Eqs.~(\ref{5})-(\ref{7}). These results can be useful in determining the parameters of the spectrum. Under conditions (\ref{9}) and (\ref{10}), i.e., at $p_x=p_x^W$, $p_z=p_z^W$, formula (\ref{5}) can be rewritten as follows:
 \begin{eqnarray}\label{17}
\varepsilon_{\nu,\mu}(p_y)-\bar\varepsilon_0^W=\nu\tilde m+\mu\sqrt{ t_2^2p_y^2 +(\frac{m_4}{m_6}\tilde m-\nu m_2)^2},~~~
\end{eqnarray}
where the indices $\mu,\nu=\pm 1$ mark the four bands, and we have introduced the designation:
 \[
  \tilde m\equiv \sqrt{m_1^2+m_5^2+m_6^2}.
 \]
Expression (\ref{17}) explicitly describes the dispersion of the four bands along the straight line perpendicular to the mirror-reflection plane and connecting the couple of the Weyl points in the Brillouin zone (Fig.~\ref{fig2}). Setting $p_y=0$ in Eq.~(\ref{17}), we find the energies of the four bands in this plane at the point $p_x=p_x^W$, $p_z=p_z^W$. With Eq.~(\ref{17}), one can also calculate the slopes $v_y^W\equiv d\varepsilon_{\nu,\mu}/dp_y$ of the two crossing bands at the Weyl points,
 \begin{eqnarray}\label{18}
v_y^{W}= \frac{t_2^2p_y^{W}}{\varepsilon^{W}-\bar\varepsilon_0^W\pm \tilde m}.
 \end{eqnarray}
Note that the combination $v_y^{W}p_y^{W}$ depends only on $m_i$.

Equations (\ref{4})-(\ref{7}) make it also possible to derive the explicit formulas for the dispersion of the two crossing bands in the immediate vicinity of the Weyl point (Appendix \ref{A}). With these formulas, the band slopes at the Weyl points along the $p_x$ and $p_z$ directions can be calculated.

If the $m_i$ ant $t_2$ were independent of $p_{\|}$, and their values satisfied Eqs.~(\ref{10}) and (\ref{14}), we would have a band-contact line in the plane $p_y=p_y^W$ rather than the Weyl point. Thus, in order to describe the dispersion of the four bands near the Weyl points, it is necessary to take into account a dependence of the parameters $m_i$ and $t_2$ on $p_{\|}$. In the vicinity of the Weyl points, these $m_i(p_{\|})$ and $t_2(p_{\|})$ can be considered as  linear functions, and the model of the spectrum has to contain the additional parameters $dm_i/dp_{\|}$ and  $dt_2/dp_{\|}$. An inspection of equations (\ref{5})-(\ref{7}) shows that they depend on $m_1$, $m_2$ and the three combinations of $m_3$, $m_4$, $m_5$, $m_6$:
\begin{eqnarray}\label{19}
\tilde\kappa&\equiv & m_5^2+m_6^2, \\
\kappa_{\perp}\equiv  m_4m_6&-&m_3m_5,\ \ \  \kappa_{\|}\equiv m_3m_6+m_4m_5,
\nonumber
 \end{eqnarray}
since $m_3^2+m_4^2=(\kappa_{\perp}^2+\kappa_{\|}^2)/\tilde\kappa$. Near the Weyl points, we will describe the linear dependences of these combinations and  $m_1$, $m_2$, $t_2$ as follows:
 \begin{eqnarray}\label{20}
 m_1(p_{\|})&=&m_1+v_1p_{\|}, \ \ m_2(p_{\|})=m_2+v_2p_{\|}, \nonumber \\
 \tilde\kappa(p_{\|})&=&m_5^2+m_6^2+\sqrt{m_5^2+m_6^2}\,\tilde vp_{\|},  \\ \kappa_{\perp}(p_{\|})&=&\frac{m_4}{m_6}(m_5^2+m_6^2)+  \sqrt{m_5^2+m_6^2}\,v_{\perp}p_{\|}, \nonumber \\
 \kappa_{\|}(p_{\|})&=&\sqrt{m_5^2+m_6^2}\,v_{\|}p_{\|},\ \
 t_2(p_{\|})=t_2+t_2'p_{\|}, \nonumber
 \end{eqnarray}
where $p_{\|}$ is reckoned from $p_{\|}^W$, $t_2'\equiv dt_2/dp_{\|}$ and the velocities $v_1$, $v_2$, $\tilde v$, $v_{\perp}$, $v_{\|}$ are the above-mentioned  additional parameters of the model spectrum, $t_2$ and all the $m_i$ in the right hand sides of these expressions are now considered as the values of these parameters at the Weyl point [i.e., they satisfy relation (\ref{10}) and are independent of $p_{\|}$], the constant $\sqrt{m_5^2+m_6^2}$ has been introduced into the formulas for $\tilde\kappa$, $\kappa_{\perp}$, $\kappa_{\|}$  to provide the necessary dimension of these quantities, and the first term in $\kappa_{\perp}(p_{\|})$ is  $\kappa_{\perp}(0)$ that is rewritten with formula (\ref{10}). Therefore, to calculate the dispersion of the bands near the two close Weyl points along an arbitrary direction of the quasimomentum, it is sufficient to express the $m_i$ ($i=3-6$) in Eqs.~(\ref{5})-(\ref{7}) through $\tilde\kappa$, $\kappa_{\perp}$, $\kappa_{\|}$ and apply formulas (\ref{20}) to the obtained expressions. Eventually, we arrive at the equation,
 \begin{eqnarray}\label{21}
 \Big[t_2^2p_y^2\!-\epsilon^2\!+d_z^2-m_1^2\! +m_2^2\!+\!\frac{\kappa_{\perp}^2+\kappa_{\|}^2}{\tilde\kappa}- \tilde\kappa\Big]^2\!\!\! -\!Y\!=0,~~~
 \end{eqnarray}
where
 \begin{eqnarray}\label{22}
&Y&=4(m_1^2\!+\!\tilde\kappa)\!\!\left[\epsilon+\frac{m_2(d_zm_1 -\kappa_{\perp})}{m_1^2+\tilde\kappa}\right]^2  \\ &-&4(m_1^2\!+\tilde\kappa-\!m_2^2) \Big[\frac{\kappa_{\|}^2}{\tilde\kappa}  +\!\frac{\tilde\kappa}{(m_1^2\!+\tilde\kappa)}\Big(\!d_z\! +\!\frac{m_1\kappa_{\perp}}{\tilde\kappa}\Big)^2\Big],~~\nonumber
\end{eqnarray}
and $m_1$, $m_2$, $\tilde\kappa$, $\kappa_{\perp}$, $\kappa_{\|}$, $t_2$ are the functions of $p_{\|}$ described by Eqs.~(\ref{20}).
It is also convenient to go from the curvilinear coordinates $p_{\perp}$ and  $p_{\|}$ to the usual Cartesian coordinates $p_{\perp}^{\rm C}$, $p_{\|}^{\rm C}$, the origin of which is at the point ($p_{\perp}=0$, $p_{\|}=p_{\|}^W$) of the band-contact line (Appendix \ref{A}). This coordinate transformation changes only the form of Eqs.~(\ref{4}) (since we use only the Cartesian coordinates below, for brevity we will omit the index ${\rm C}$ in $p_{\|}^{\rm C}$,  $p_{\|}^{\rm C}$),
\begin{eqnarray}\label{23}
d_z&= &a'p_{\perp}+\frac{a'p_{\|}^2}{2R}, \\
\bar\varepsilon_0&= &\varepsilon_0(0)+\frac{d\varepsilon_0}{dp_{\|}}p_{\|}+ap_{\perp}+ \frac{1}{2}\frac{d^2\varepsilon_0}{dp_{\|}^2}p_{\|}^2+\frac{ap_{\|}^2}{2R},~~~ \nonumber
\end{eqnarray}
where $R$ is the radius of the curvature for the band-contact line  at the point $p_{\|}=0$, and we have expanded $\varepsilon_0(p_{\|})$ in powers of $p_{\|}$.
[The last two terms in the formula for $\bar\varepsilon_0$ can be  neglected for the W2 points, but these terms are important for the W1 points since $(d\varepsilon_0/dp_{\|})=0$ in this case.]
Equations (\ref{20})-(\ref{23}) together with the definition $\varepsilon=\epsilon+\bar\varepsilon_0$  completely define the model spectrum proposed in this paper.

In Appendix \ref{B}, we discuss how $t_2'$ and the velocities $v_1$, $v_2$, $\tilde v$, $v_{\perp}$, $v_{\|}$ can be found from the band-structure calculations along the $p_{\|}$ direction near the Weyl points. However, such calculation are rare \cite{wu}. In this context, it is also shown in Appendix \ref{B} that for the  not-too-strong spin-orbit interaction, a minimal model can be formulated that takes into account the main feature of the Weyl spectrum, the linear splitting of the crossing bands in the immediate vicinity of the Weyl point. Within this model, we set
 \begin{eqnarray}\label{24}
v_1=v_2=\tilde v=v_{\perp}=t_2'=0,\ \ \ v_{\|}\neq 0,
  \end{eqnarray}
and the only nonzero velocity $v_{\|}$  determines the linear splitting of the bands in the $p_{\|}$ direction.

Below, using NbP as an example, we will consider the W1 and W2 points in the TaAs family of the topological semimetals in more detail.

\section{Charge-carrier spectrum for ${\rm NbP}$}\label{NbP}

Let us apply the model spectrum described in the previous section to NbP. To find the values of the parameters of the model in this case, we compare the results mentioned in Sec.~\ref{spectrum} with appropriate data obtained in the numerical band-structure calculations  \cite{lee} for NbP.

\subsection{W1 points}\label{W1}

\begin{table}[t]
\caption{\label{tab1} Quantities characterizing the W1 points in NbP. The values of these quantities were obtained from numerical calculations \cite{lee} of the electron-band structure of NbP  near the W1 point marked in Fig.~\ref{fig1}. All the energies are reckoned from the Fermi level, $\varepsilon_i(0)$ are the energies of the bands at the point $(p_x^{W1},0,2\pi\hbar/c)$ of the mirror-reflection plane $p_y=0$ (only $\varepsilon_i(0)$ for $i=2,3,4$  can be extracted from Fig.~8a of Ref.~\cite{lee}). Here $a=3.334$ \AA \ \cite{lee}. }
\begin{tabular}{|cccccc|c|}
\hline
\hline \\[-2.5mm]
$\varepsilon_2(0)$&$\varepsilon_3(0)$&$\varepsilon_4(0)$&$\varepsilon^{W1}$&$p_y^{W1}$& $v_{y\pm}^{W1}$&$v_{x\pm}^{W1}$ \\
 meV &meV &meV&meV&$\frac{2\pi\hbar}{a}$ &$10^5\frac{m}{s}$&$10^5\frac{m}{s}$   \\
 \colrule
-60.3&-37.4&-26&-53.1&0.0028&3.7&1.5   \\
 & & & & &-5.7 &-3  \\
\hline \hline
\end{tabular}
\end{table}

\begin{table}[t]
\caption{\label{tab2}  Values of the parameters  for the W1 points of NbP. For these points, $m_3=m_5=0$. The values of the other parameters  are obtained, using  the data presented in Table  \ref{tab1} (see the text). Here  $\tilde m\equiv \sqrt{m_1^2+m_6^2}$.}
\begin{tabular}{|ccccc|cccc|}
\hline
\hline \\[-2.5mm]
$m_2$&$\tilde m$&$(m_4/m_6)$&$\bar\varepsilon_0^{W1}$ &$t_2$&$m_1$& $a$&$a'$&$v_{\|}$ \\
 meV &meV& &meV&$10^5\frac{m}{s}$&meV &$10^5\frac{m}{s}$&$10^5\frac{m}{s}$&$10^5\frac{m}{s}$ \\
 \colrule
15.7 &27.15&0.368 &-58.86 &5.92 &$\approx 0$\footnote{This value of $m_1$ leads to $m_6\approx 27.15$ meV, $m_4\approx 10$ meV.} &-0.75 &2.7&0.095   \\
\hline \hline
\end{tabular}
\end{table}

Consider a pair of the close W1 points, position of which in the Brillouin zone of NbP  ($p_x^{W1},\pm p_y^{W1},2\pi\hbar/c$) is indicated in Fig.~\ref{fig1}. As was mentioned above,  equations (\ref{8}) give $m_3=m_5=0$, and condition (\ref{10}) is fulfilled due to the symmetry. To estimate the parameters of the spectrum near the W1 points, we may use the following results: (i) At the point ($p_x^{W1},0,2\pi\hbar/c$), the energies $\varepsilon_i(0)$ of the bands are described by formula (\ref{17}). (ii) The energy of the Weyl point is given by Eq.~(\ref{12}). (iii) The value of $p_y^{W1}$ is determined by Eq.~(\ref{14}). (iv) The  slopes of the two crossing bands at the W1 point along the $p_y$ direction are described by expression (\ref{18}). The values of all these quantities found in the band-structure calculations \cite{lee} are presented in Table \ref{tab1}. Therefore, we have seven relationships which turn out to  depend on the five  parameters: $t_2$, $\bar\varepsilon_0^{W1}$, $\tilde m^2\equiv m_6^2+m_1^2$, $m_4/m_6$, and $m_2$, and so we are able not only to find values of these parameters (Table \ref{tab2}), but also to check them; see Fig.~\ref{fig2}.

\begin{figure}[t] % %%%%%%%%%%%%%%%%%%%%%%%%%%%%%%%%%%%%%
 \centering  \vspace{+9 pt}
\includegraphics[scale=0.44]{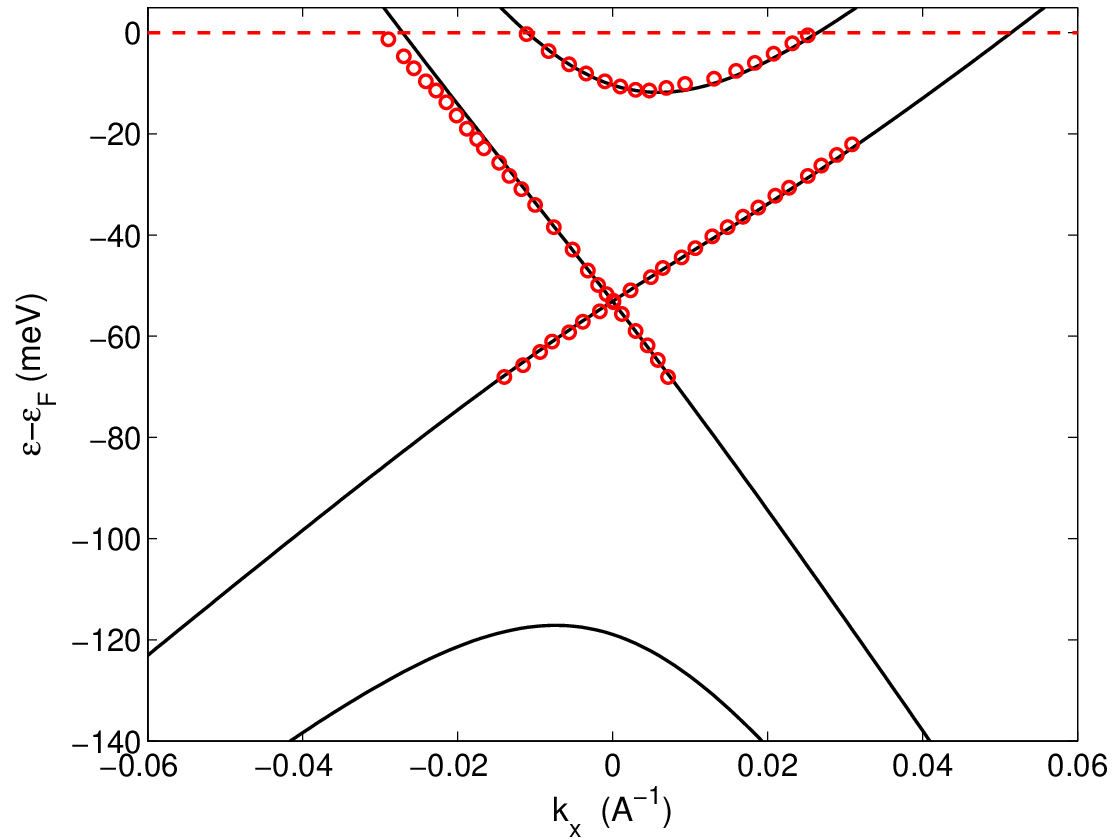}
\caption{\label{fig3} Dispersion of the bands along the $p_x$ axis ($k_x=p_x/\hbar$) at $p_z=2\pi\hbar/c$, $p_y=p_y^{W1}$ for the parameters from Table \ref{tab2}. The solid lines are plotted with Eqs.~(\ref{5})--(\ref{7}). The $p_x$ is measured from $p_x^{W1}$. The red cycles are the data of Fig.~8b in Ref.~\cite{lee} for the W1 point in NbP (see Fig.~\ref{fig1}).
}
\end{figure}

At a given value of $m_1$, equations (\ref{a6}), (\ref{a10}) provide possibility of calculating the slopes of the two crossing bands at the Weyl point along the $p_x$ direction (i.e., along $p_{\perp}$ at $p_{\|}=0$), and  therefore, they make it possible to determine the parameters $a'$ and $a$, using the values of $v_{x\pm}^{W1}$ from Table \ref{tab1} \cite{com2}. The parameter $m_1$ can be found, using a fit of the dispersion of the fourth band along the $p_x$ direction at $p_y=p_y^{W1}$, $p_z=2\pi\hbar/c$  to the  results of the numerical calculations in Ref.~\cite{lee} (Fig.~\ref{fig3} and Table \ref{tab2}).

\begin{figure}[t] % %%%%%%%%%%%%%%%%%%%%%%%%%%%%%%%%%%%%%
 \centering  \vspace{+9 pt}
\includegraphics[scale=0.44]{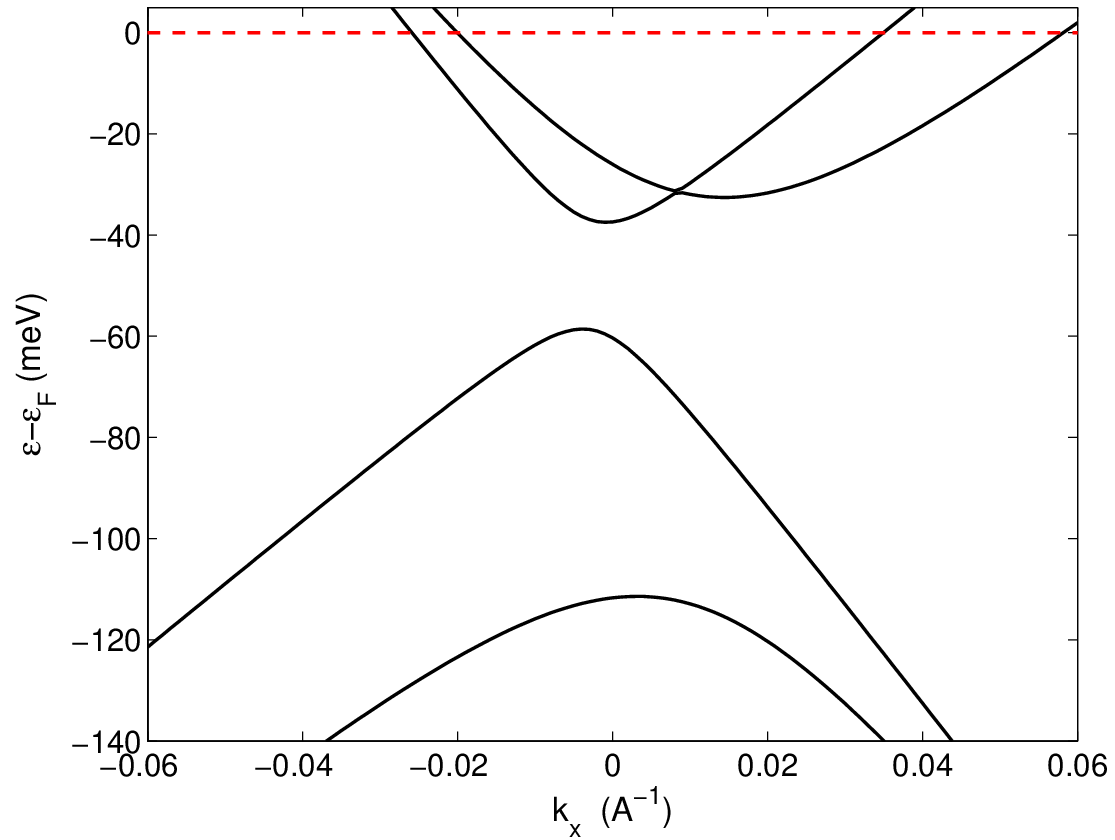}
\caption{\label{fig4}
Dispersion of the bands along the $p_x$ axis at $p_z=2\pi\hbar/c$, $p_y=0$ (i.e., along the $Z-S$ axis of the Brillouin zone) for $\tilde m$, $\bar\varepsilon_0$, $t_2$ from Table \ref{tab2} but with $m_2=10$ meV and $m_4/m_6=0.578$ (the other parameters $m_1=9$ meV, $a=-0.54\times 10^5$ m/s, $a'=2.46\times 10^5$ m/s, are found from a fit like in Fig.~\ref{fig3}). The lines are plotted with Eqs.~(\ref{5})--(\ref{7}). The $p_x$ is measured from $p_x^{W1}$. Note that the additional Weyl point is visible in the plane $p_y=0$.
}
\end{figure}

It is also necessary to emphasize that Table \ref{tab1} leads to two sets of the parameters that give the same dispersion of the bands along $p_y$ direction (see the caption to Figs.~\ref{fig2}). One of the two sets is obtained under condition (\ref{16}), and only this set is presented in Table \ref{tab2}. The other possible set corresponds to the opposite inequality in  relation (\ref{16}). However, for this set, at any value of $m_1$, apart from the well-known W1 points, additional Weyl points exist in the energy spectrum in the $p_x$ axis (i.e., in the $Z-S$ axis of the Brillouin zone). One of these points always lies  in the vicinity of $p_x^{W1}$ (the other Weyl point can be far away from $p_x^{W1}$), Fig.~\ref{fig4}. For the first set of the parameters, a similar point can appear only at $|m_1|\gtrsim m_2$. However, Refs.~\cite{h.weng,lee,klotz,grassano} do not report  any Weyl point on the $p_x$ axis in the vicinity of $p_x=p_x^{W1}$. Therefore, this result permits us to exclude the second set from the consideration.

    %%%%%%%%%%%%%%%%%%%%%%%%%%%%%%%%%%%%%%%%%%

    %%%%%%%%%%%%%%%%%%%%

In the case of the W1 point, due to the symmetry relations (\ref{8}), the linear dependence of $m_i$ on $p_{\|}$ occurs only for $m_3(p_{\|})$ and $m_5(p_{\|})$. For the other $m_i$ and for $t_2$, one has $m_i(p_{\|})-m_i(0)\propto p_{\|}^2$, $t_2(p_{\|})-t_2(0)\propto p_{\|}^2$, and we may neglect these weak dependences. Then, $v_1=v_2=\tilde v=v_{\perp}=t_2'=0$ in formulas (\ref{20}), and the minimal model (\ref{24}) is well applicable to the W1 point. Besides,  the coefficient $(d\varepsilon_0/dp_{\|})$ in Eqs.~(\ref{23}) also  vanishes due to the symmetry.

 \begin{figure}[t] % %%%%%%%%%%%%%%%%%%%%%%%%%%%%%%%%%%%%%
 \centering  \vspace{+9 pt}
\includegraphics[scale=0.44]{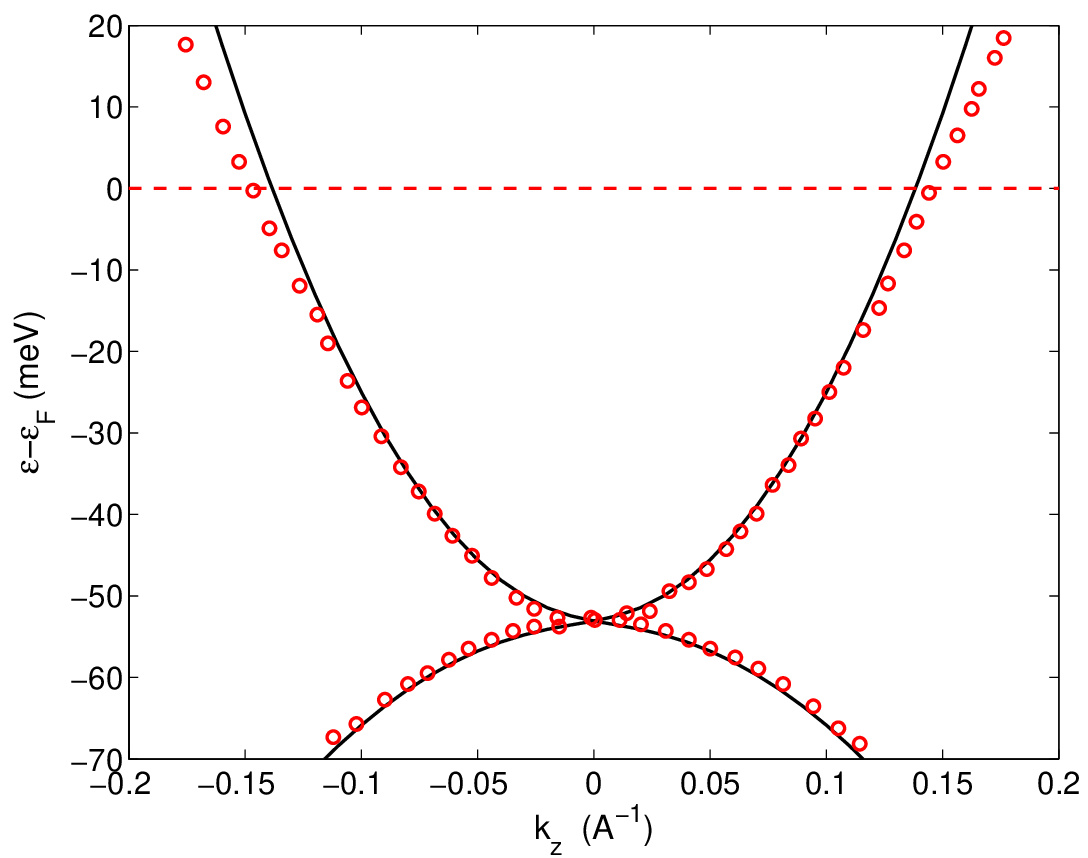}
\caption{\label{fig5} Dispersion of the two crossing bands along the $p_z$ direction ($k_z=p_z/\hbar$) in NbP at $p_x=p_x^{W1}$, $p_y=p_y^{W1}$, Eq.~(\ref{25}).  The black solid lines show the bands for the parameters presented in Table \ref{tab2} and for  $R=0.2(2\pi\hbar/a)$, $d^2\varepsilon_0/dp_{\|}^2\approx 0.37/m$, and $v_{\|,\|}=0.79\times 10^4$ m/s (which corresponds to $v_{\|}=0.95\times 10^4$ m/s)  where $a=3.334$ \AA,  and $m$ is the free-electron mass. Note that the black solid lines remain practically unchanged if $v_{\|,\|}< 10^4$ m/s. The red cycles are the data of Fig.~8c in Ref.~\cite{lee}. These data were obtained for the W1 point marked in Fig.~\ref{fig1}. All the energies are measured from the Fermi level (red dashed line).
}
\end{figure}   %%%%%%%%%%%%%%%%%%%%%%%%%%%%%%%%%%%%%%%%%%

To describe the dispersion of the two crossing bands near the W1 point along the $p_z$ axis (i.e., along the $p_{\|}$ direction), we set $\Delta p_{\perp}=0$ in Eqs.~(\ref{a3}), (\ref{a6}), (\ref{a10}) and take into account that $m_1\approx 0$ (Table \ref{tab2}) and $(d\varepsilon_0/dp_{\|})=0$ in Eqs.~(\ref{a6}) and (\ref{a10}), respectively. Then, we obtain
\begin{eqnarray}\label{25}
\Delta\varepsilon=
\frac{1}{2}\frac{d^2\varepsilon_0}{dp_{\|}^2}p_{\|}^2+ \frac{ap_{\|}^2}{2R}
\pm \sqrt{\frac{q_2(a')^2p_{\|}^4}{4q_1^2R^2}+ (v_{\|,\|}p_{\|})^2},~~
 \end{eqnarray}
where $q_1$ and $q_2$ are defined by Eqs.~(\ref{a7}), $R$ is the radius of the curvature of the band-contact ring near the W1 point, and the velocity $v_{\|,\|}$ is related to $v_{\|}$ by formula (\ref{b2}). Thus, the dispersion of the crossing bands in the $p_{\|}$ direction is a superposition of the quadratic and linear terms, and the quadratic terms in Eq.~(\ref{25}) are not negligible since the velocity $v_{\|,\|}$  may be relatively small. We fit the bands described by Eq.~(\ref{25}) to the appropriate data of Lee {\it et al}.\ \cite{lee} (Fig.~\ref{fig5}). The fit leads to the conclusion that $v_{\|,\|}$ is indeed small, $ v_{\|,\|}\lesssim 10^4$ m/s. We assume below that $v_{\|,\|}$ is approximately equal to $0.79\times 10^4$ m/s, the value found in the band-structure calculations of Grassano {\it et al}. \cite{grassano}. This assumption regarding $v_{\|,\|}$ is justified by the fact that in Ref.~\cite{grassano}, the other slopes of the bands (i.e., the slopes along the $p_x$ and $p_y$ directions) near the W1 point are close to the values of $v_{x\pm}^{W1}$, $v_{y\pm}^{W1}$ in Table \ref{tab1}.
According to formula (\ref{b2}), the chosen value of $v_{\|,\|}$ corresponds to $v_{\|}\approx 0.95\times 10^4$ m/s.
From the fit, we also obtain $d^2\varepsilon_0/dp_{\|}^2\approx 0.37/m$  ($m$ is the free-electron mass) and $R\approx 0.2$ in units of $2\pi\hbar /a$ where $a=3.334$\AA. The obtained $R$ is comparable with $0.15(2\pi\hbar /a)$, the mean radius of  the band-contact ring calculated in Ref.~\cite{lee}.

The obtained values of the parameters can be verified, using the quantum-oscillation frequencies associated with extremal cross sections of the Fermi surface. Without any additional fit, for the parameters presented in Table \ref{tab2}, formulas of Appendix \ref{C} predict the frequencies  $F_{\alpha1}\approx 28.6$ T, $F_{\alpha2}\approx 6.5$ T, generated by the electrons in the bands $\varepsilon_3({\bf p})$ and $\varepsilon_4({\bf p})$, respectively, if the magnetic field $H$ is parallel to the $p_z$ axis. These frequencies are slightly less than the experimental values measured by Klotz {\it et al}.\ \cite{klotz}: $F_{\alpha1}=32.1$ T, $F_{\alpha2}=8.7$ T. However, the difference  $\varepsilon_F- \varepsilon^{W1}=57$ meV in Ref.~\cite{klotz} is 4 meV larger than the difference of the same energies in the work of Lee {\it et al}.\ \cite{lee} (see Table \ref{tab1}). The $4$ meV upward shift of the Fermi level in Table \ref{tab2} leads to a better agreement between the calculated frequencies  ($F_{\alpha1}\approx 32$ T, $F_{\alpha2} \approx 8.14$ T) and the experimental data.

In verifying the values of the parameters, the angular dependences of the  extremal cross sections can be also useful \cite{m-sh21a,m24}. Due to  relatively small values of $v_{\|}$ for the semimetals of the TaAs family [and $(d\varepsilon_0/dp_{\|})_{W1}=0$], the Fermi-surface pockets surrounding the W1 points are significantly elongated along the nodal rings. However, in the case of TaAs, for which the spin-orbit interaction has the largest magnitude, each of the W1 points is enclosed by a separate Fermi-surface pocket, the elongation of which in the $p_{\|}$ direction is noticeably less than in the other semimetals of this family. Then, the model of Sec.~\ref{spectrum} is expected to be applicable for calculating the cross-sectional areas of the pocket  for all directions of  $H$, including $H\parallel$ $p_x$, $p_y$. (The angular dependences of $F_{\alpha1}$ for TaAs  were measured in Ref.~\cite{Arnold}.) For other semimetals of TaAs family, the electron pocket surrounding both the close W1 points is more elongated along the nodal ring than in TaAs. Then, expansions (\ref{20}), which take into account only constant terms and terms linear in $p_{\|}$, may become insufficient for the accurate calculation of the cross-sectional areas of the pockets when the angle between the magnetic field and the $p_z$ axis is close to $\pi/2$. In this situation, the terms proportional to $p_{\|}^2$ should be added to the expansions (\ref{20}) \cite{comm}. The additional parameters appearing in such expansions can be found from the angular dependences of the  quantum-oscillations frequencies.

\subsection{W2 points}\label{W2}

\begin{table}[t]
\caption{\label{tab3} Quantities characterizing the W2 points in NbP. Values of these quantities were found with the band-structure calculations in Ref.~\cite{lee} for the W2 point marked in Fig.~\ref{fig1}. At this point, $\theta\approx -127^{\circ}$. For comparison, the quantities calculated in Ref.~\cite{klotz} are also shown.  Note that although $\varepsilon_2(0)$, $\varepsilon_3(0)$, $\varepsilon^{W2}$ differ  significantly in Refs.~\cite{lee} and \cite{klotz}, the differences $\varepsilon_3(0)-\varepsilon_2(0)$ and $\varepsilon^{W2}-\varepsilon_2(0)$  are close to each other, i.e., the bands are shifted as a whole in these works.
}
\begin{tabular}{|c|ccccccc|cc|}
\hline
\hline \\[-2.5mm]
Ref.&$\varepsilon_1(0)$&$\varepsilon_2(0)$&$\varepsilon_3(0)$&$\varepsilon_4(0)$&
$\varepsilon^{W2}$&\!\!\!$p_y^{W2}$&\!\!\! $v_{y\pm}^{W2}$&$v_{x\pm}^{W2}$&$v_{z\pm}^{W2}$ \\
 &meV&meV &meV &meV&meV&$\frac{2\pi\hbar}{a}$ &$10^5\frac{m}{s}$&$10^5\frac{m}{s}$&$10^5\frac{m}{s}$  \\
 \colrule
\cite{lee} & & 19.2&38.1 & &26.1 &0.0049&2.1 &2.1 &3.8 \\
 & & & & & & &-3.2 &-1.6 & -1 \\
\colrule
\cite{klotz}&-68 &-1.6 &18 &49 &5 &  & & & \\
\hline \hline
\end{tabular}
\end{table}

\begin{table}[t]
\caption{\label{tab4}  Values of the parameters  for the W2 points of NbP. These values are obtained, using  the data \cite{lee} presented in  Table \ref{tab3}. The parameter $m_1$ is found from the fit of the bands calculated with the minimal model (see the text) to the data of Ref.~\cite{lee} shown in Figs.~\ref{fig7} and \ref{fig8}. Here $\theta=-127.3^{\circ}$, and $\tilde m\equiv \sqrt{m_1^2+ m_5^2+m_6^2}$. The value of $v_{\|}$ corresponds to $v_{\|,\|}=0.44\times 10^5$ m/s.}
\begin{tabular}{|ccccc|ccccc|}
\hline
\hline \\[-2.5mm]
$m_2$&$\tilde m$&\!\!$\frac{m_4}{m_6}$&$\bar\varepsilon_0^{W2}$ &$t_2$&$m_1$&$a$&$a'$&$\frac{d\varepsilon_0}{dp_{\|}}$&$v_{\|}$ \\
meV&meV&\!\! &\!meV&\!$10^5\frac{m}{s}$& meV&\!$10^5\frac{m}{s}$&\!$10^5\frac{m}{s}$ &\!$10^5\frac{m}{s}$&\!$10^5\frac{m}{s}$
\\
 \colrule
30.5 &40&\!\!\!0.272 &\!17.8 &\!4.1 &\!-8&\!-1.95&\!4.6 &\!-0.65 &\!0.67  \\
\hline \hline
\end{tabular}
\end{table}

\begin{figure}[t] % %%%%%%%%%%%%%%%%%%%%%%%%%%%%%%%%%%%%%
 \centering  \vspace{+9 pt}
\includegraphics[scale=0.44]{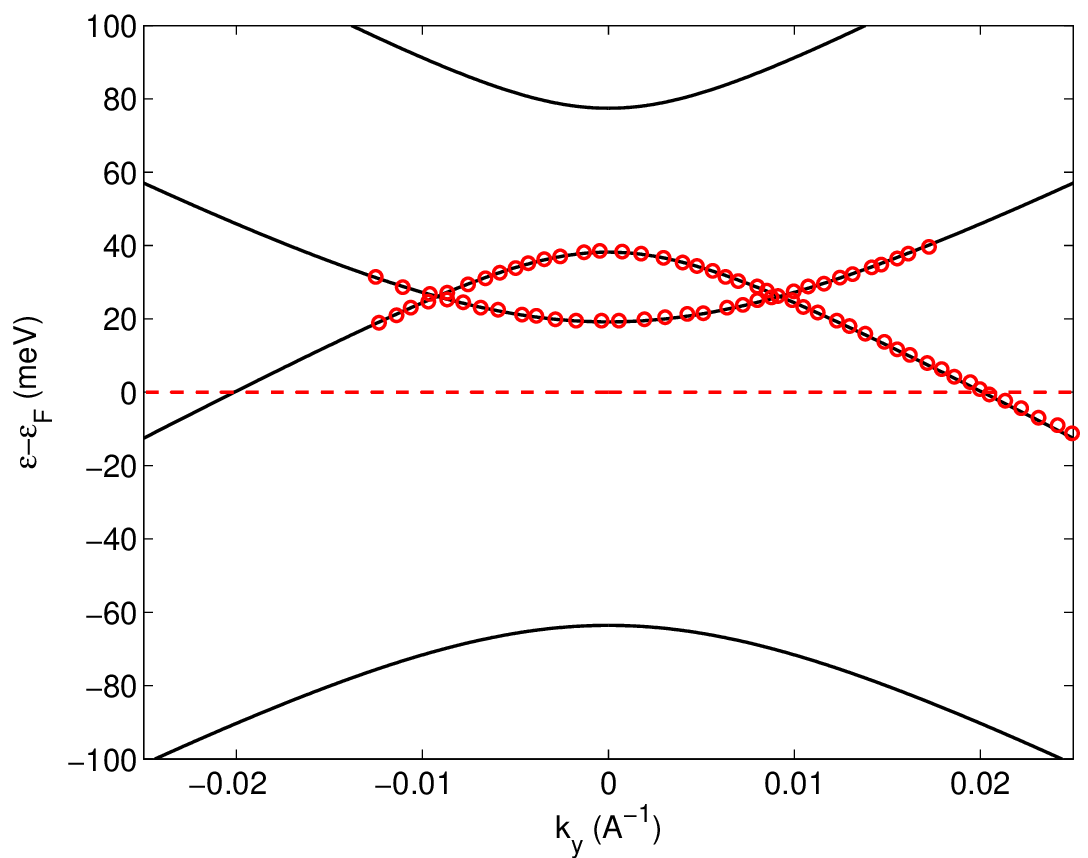}
\caption{\label{fig6} Dispersion of the four bands along the $p_y$ direction ($k_y=p_y/\hbar$) in NbP at $p_x=p_x^{W2}$, $p_z=p_z^{W2}$, Eq.~(\ref{17}). The point ($p_x^{W2}$,$p_z^{W2}$) in the mirror-reflection plane $p_y=0$ is the projection  of the two W2 points onto this  plane. The bands are plotted for the  parameters presented in Table \ref{tab4}. The red cycles are the data of Fig.~9a in Ref.~\cite{lee} for the W2 point in NbP (Fig.~\ref{fig1}). The energies  are measured from the Fermi level (red dashed line).
}
\end{figure}

Consider a pair of the close W2 points in NbP (this pair is marked in Fig.~\ref{fig1}). Condition (\ref{10}) means that $m_3=-(m_4/m_6)m_5$ near these points. Like for the W1 points, the dispersion of the bands along the $p_y$ axis, Eq.~(\ref{17}), is determined by the five parameters: $t_2$, $\bar\varepsilon^{W2}$,  $m_2$, $m_4/m_6$, and $\tilde m\equiv \sqrt{m_1^2+m_5^2+m_6^2}$. Using the values of the quantities presented in  Table \ref{tab3}, these parameters are found (Table \ref{tab4}), see also Fig.~\ref{fig6}. As in the case of the W1 points, we find two sets of the parameters for the data of Table \ref{tab3}. The first set is obtained under  condition (\ref{16}),
whereas the second set is derived at the opposite inequality. Both the sets  describe the dispersion of the bands along the $p_y$ direction equally well. As in the case of the W1 points, the second set of the parameters  leads to the appearance of additional Weyl points in the mirror-reflection plane. Such points have never been detected in the band-structure calculations and experiments, and so we exclude the second sets of the parameters from our consideration.

    %%%%%%%%%%%%%%%%%%%%%%%%%%%%%%%%%%%%%%%%%%

Let $\theta$ be the angle between the direction of the $p_{\|}$ axis at the W2 point (i.e., the direction of the tangent to the band-contact line at this point)  and the $p_z$ axis. Our analysis of the band-contact ring calculated in Ref.~\cite{lee} gives $-124^{\circ} \gtrsim \theta \gtrsim -128^{\circ}$ for the W2 point marked in Fig.~\ref{fig1}. (This definition of $\theta$ follows from the requirement of a continuous change of $\theta$ along the line and from $\theta=0$ for the W1 point,  Fig.~\ref{fig1}.)
Below we will consider $|\theta|$ lying in the interval $124^{\circ}$--$128^{\circ}$. Note that the calculations of Wu {\it et al}. \cite{wu} give the value $|\theta|\approx 134^{\circ}$, which is not far away from this interval \cite{comm-wu}.

The results of Sec.~\ref{W1} show that in NbP, the linear splitting of bands at the W1 point along the $p_{\|}$ direction  is relatively small. Therefore, one can  expect that the same is true for the W2 point. Then, the simple minimal model defined in Sec.~\ref{spectrum} and Appendix \ref{B}  can be used to describe the four bands near the W2 point. Within this model, if values of $m_1$ and $\theta$ are fixed, formulas (\ref{b2}) and (\ref{b3}) determine values of $a'$, $a$, $v_{\|,\|}$, $d\varepsilon_0/dp_{\|}$, and $v_{\|}$ in terms of $v_{x\pm}^{W2}$  and $v_{z\pm}^{W2}$ presented in Table \ref{tab3}.

 \begin{figure}[t] % %%%%%%%%%%%%%%%%%%%%%%%%%%%%%%%%%%%%%
 \centering  \vspace{+9 pt}
\includegraphics[scale=0.44]{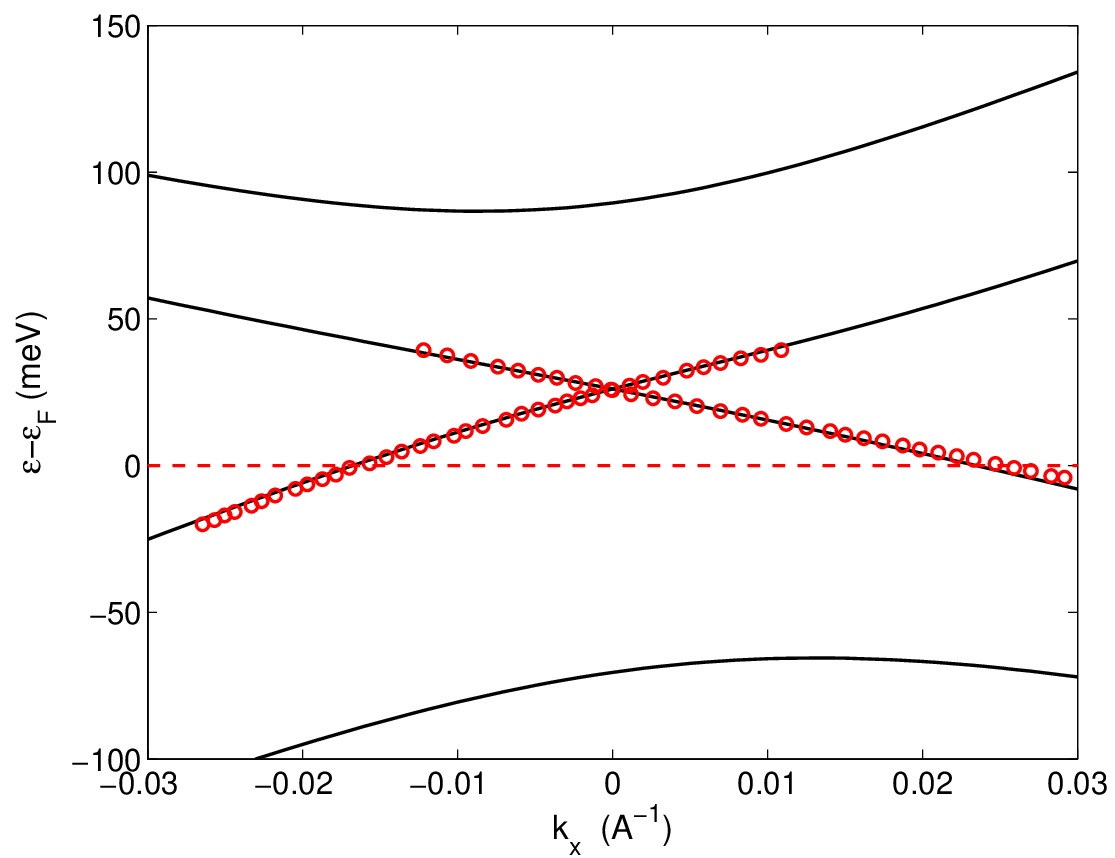}
\caption{\label{fig7} Dispersion of the four bands along the $p_x$ direction ($k_x=p_x/\hbar$) in NbP at $p_y=p_y^{W2}$, $p_z=p_z^{W2}$. The bands are plotted with Eqs.~(\ref{20})--(\ref{24}) for the set  of the  parameters in Table \ref{tab4} and $\theta=-127.3^{\circ}$.
The red cycles are the data of Fig.~9b in Ref.~\cite{lee} for the W2 point in NbP (Fig.~\ref{fig1}). The energies  are measured from the Fermi level (red dashed line).
}
\end{figure}   %%%%%%%%%%%%%%%%%%%%%%%%%%%%%%%%%%%%%%%%%%

 \begin{figure}[t] % %%%%%%%%%%%%%%%%%%%%%%%%%%%%%%%%%%%%%
 \centering  \vspace{+9 pt}
\includegraphics[scale=0.44]{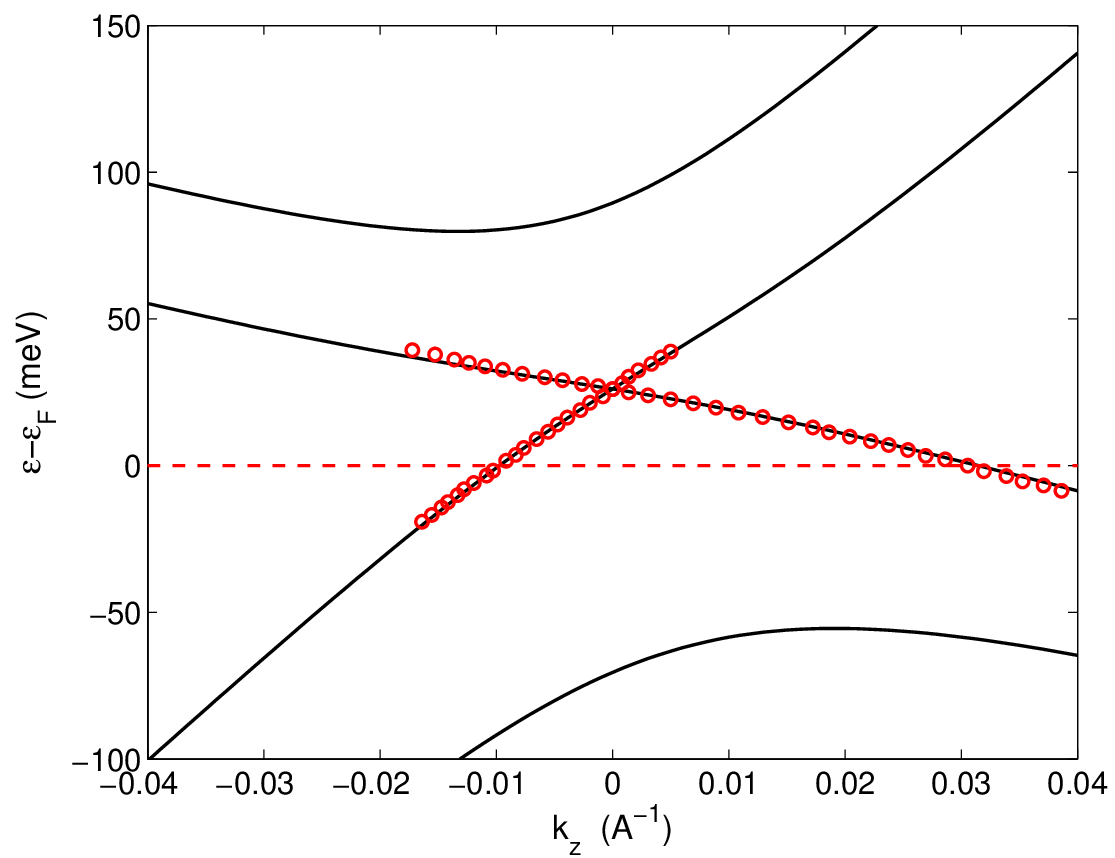}
\caption{\label{fig8} Dispersion of the four bands along the $p_z$ direction ($k_z=p_z/\hbar$) in NbP at $p_y=p_y^{W2}$, $p_x=p_x^{W2}$. The bands are plotted with Eqs.~(\ref{20})--(\ref{24}) for the  set  of the  parameters in Table \ref{tab4} and $\theta=-127.3^{\circ}$.
The red cycles are the data of Fig.~9c in Ref.~\cite{lee} for the W2 point in NbP (Fig.~\ref{fig1}). The energies are measured from the Fermi level (red dashed line).
}
\end{figure}   %%%%%%%%%%%%%%%%%%%%%%%%%%%%%%%%%%%%%%%%%%

Figures \ref{fig7} and \ref{fig8} show the dispersion of the bands along the $p_x$ and $p_z$ directions. These bands are calculated with Eqs.~(\ref{20})-(\ref{23}), using the minimal model Eq.~(\ref{24}), and the values of parameters from Table \ref{tab4}. The value of $m_1\approx -8$ meV is found with the best fit  of the calculated  second and third bands to the data of Ref.~\cite{lee}. Interestingly, our calculations show that with changing the value of  $m_1$, the positions of the maximum of the first band and of the minimum of the fourth band shift in opposite directions. Therefore, the appropriate information on these bands could verify the obtained value of $m_1$.
We also find that the terms proportional to  $p_{\|}^2$ in Eqs.~(\ref{23}) have a small effect on the presented plots since the $p_{\|}$ direction noticeably deviates from the $p_x$ and $p_z$ axes in the case of the W2 points. We omit these terms when calculating the plots in Figs.~\ref{fig7} and \ref{fig8}.

As explained in Appendix \ref{B}, for the applicability of the minimal model, the angle $|\theta|$ has to be close to its critical value $\theta_0$ determined by Eq.~(\ref{b4}).  This is indeed the case for the W2 point in NbP since $|\theta_0|\approx 128^{\circ}$ for the values of $v_{x\pm}$, $v_{z\pm}$ in Table \ref{tab3}. In our calculations of the bands, we try different values of  $|\theta|$ lying in the interval from $124^{\circ}$ to $128^{\circ}$. In this case, only $v_{\|,\|}$ noticeably changes, whereas the other parameters are little affected. For example, at $\theta=-127.6^{\circ}$, $-127.3^{\circ}$ $-127^{\circ}$, $-126.5^{\circ}$, and $-125^{\circ}$, we obtain $v_{\|,\|}=0.126$, $0.44$, $0.59$, $0.77$, $1.08$ (in units of $10^5$ m/s), respectively. However, when $|\theta|$ becomes less than $126.5^{\circ}$, the quality of the fit in Figs.~\ref{fig7} and \ref{fig8} deteriorates.  Specifically, a good fit along the $p_x$ axis worsens the fit along the $p_z$ axis, and vice versa. This result supports the assumption on the applicability of the minimal model. In order to choose $\theta$ (and thus $v_{\|,\|}$), the band-structure calculation in the $p_{\|}$ direction would be desirable. Such a calculation was made by Wu {\it et al}. \cite{wu} in order to prove that the W2 point is of the type II. Our analysis of their Fig.~3c gives the estimate, $v_{\|,\|}/|d\varepsilon_0/dp_{\|,\|}| \sim 0.7$. In Table \ref{tab4}, we use  $\theta=-127.3^{\circ}$, which leads to the close value of this ratio. Interestingly, although for all $\theta$ obeying the condition  $-128^{\circ}<\theta\lesssim -126.5^{\circ}$, we obtain that the W2 point is of the type II, at the angle $\theta=-126.5^{\circ}$, we find that the W2 point is of type I since $v_{\|,\|}=0.77\times 10^5 {\rm m/s}>|d\varepsilon_0/dp_{\|}| =0.63\times 10^5$ m/s in this case. In other words, the type of the W2 points is very sensitive to the angle $\theta$ used in the calculations.

\section{Discussion}\label{discussion}

In Sec.~\ref{NbP}, we have found values of the parameters for the model spectrum near  the W1 and W2 points of NbP, using the data of the band-structure calculation \cite{lee} in the vicinities of these points. Once the values are found, the dispersion of the four bands in the region of the Weyl points and  the cross-sectional areas of the Fermi surface can be analyzed with simple equations (\ref{20})--(\ref{24}), without resorting to the complicated band-structure calculations. It is clear that this approach can be extended to NbAs, TaAs, TaP. However, the  question remains about the accuracy of the obtained parameters since different band-structure calculations can give noticeably different energies of the bands.
For example, this conclusion follows from the characteristic energies found near the W2 point of NbP in Ref.~\cite{klotz} (Table \ref{tab3}). Estimating the above-mentioned accuracy, we note that although in Table \ref{tab3}, the energies $\varepsilon_2(0)$, $\varepsilon_3(0)$, $\varepsilon^{W2}$ obtained by Klotz {\it et al}. \cite{klotz} differ  significantly from the appropriate data of Refs.~\cite{lee}, the differences of these energies, $\varepsilon_3(0)-\varepsilon_2(0)$ and $\varepsilon^{W2}- \varepsilon_2(0)$,  are close to each other. In table \ref{tab5}, we present the values of the parameters found from the accessible data of the band-structure calculations in Refs.~\cite{grassano} and \cite{klotz}. Excepting $\bar\varepsilon^{W}$, the largest deviations of the values (about 20\%) from the values in Tables \ref{tab2} and \ref{tab4} occur for the parameters $m_i$.  These deviations seem to characterize the ``sensitivity'' of the  parameters to the results of the band-structure computations. As to the parameter $\bar\varepsilon^{W}$, it is determined by the position of Fermi level in the crystal. This position depends on the total band structure of the semimetal rather than on its local parts near the Weyl points. This explains the significant difference in $\bar\varepsilon^{W2}$ for data of Refs.~\cite{lee} and \cite{klotz}. Therefore, it would be preferable to refine the parameter $\bar\varepsilon^{W}$,  using experimental data on an extremal cross section of the Fermi surface (when such a cross section exists near a Weyl point).

\begin{table}[t]
\caption{\label{tab5} Values of the parameters for the W1 and W2 points of NbP; cf. Tables \ref{tab2} and \ref{tab4}. These values are found, using the accessible data \cite{comm-table} of the band-structure calculations in Refs.~\cite{grassano} and \cite{klotz}, respectively.}
\begin{tabular}{|c|ccccc|ccc|}
\hline
\hline \\[-2.5mm]
point&$m_2$&$\tilde m$&$(m_4/m_6)$&$\bar\varepsilon_0^{W}$ &$t_2$&$m_1$& $a$&$a'$ \\
 &meV &meV& &meV&$10^5\frac{m}{s}$&meV &$10^5\frac{m}{s}$&$10^5\frac{m}{s}$ \\
 \colrule
W1&18.7 &31&0.443&-64.3&6.16 &-2&-0.78 &2.76\\
W2&24.4&32.6&0.272&-0.65& & & &\\
\hline \hline
\end{tabular}
\end{table}

The described origin of the Weyl points in the topological materials of the TaAs family points out to a possible direction for searching for other noncentrosymmetric Weyl semimetals. It is well known \cite{herring} that in neglect of the spin-orbit interaction, an accidental contact of two bands of the opposite  parities can occur along lines in the mirror-reflection planes of noncentrosymmetric crystals. For example, the inversion of these  bands at some point of the plane is sufficient for the appearance of such a line. Then, the Hamiltonian of Sec.~\ref{spectrum} can be applied to describe the energy spectrum near this nodal line. If the plane contains a  two-, four-, or six-fold rotation axis, which may be denoted as the $z$ axis, and the nodal line in this plane crosses the plane, $p_z=0$, or the Brillouin-zone faces perpendicular to $p_z$, the conditions like  Eqs.~(\ref{8}) can lead to existence of the W1 points in such a material.

Finally, let us compare the model spectrum of Sec.~\ref{spectrum}  with the model suggested by Jiang {\it et al.} \cite{jiang}  and  used in the articles \cite{zhao,moon}. In Ref.~\cite{jiang}, the following Hamiltonian was proposed to describe  a pair of the Weyl points:
\begin{eqnarray}\label{26}
\hat H=m\tau_z+b\sigma_x+v\tau_x(p_y\sigma_x+p_x\sigma_y+ p_z\sigma_z),
\end{eqnarray}
where $\tau_i$ and $\sigma_i$ are the Pauli matrices, $m$,  $b$, and $v$ are constants, and we have interchanged the  $p_x$ and $p_y$ axes in order to agree the notations with our choice of the coordinate axes. This Hamiltonian leads to the following dispersion of the four bands \cite{jiang}:
\begin{eqnarray}\label{27}
\epsilon_{s,\mu}&=&s\Big[m^2+(vp_x)^2+(vp_y)^2+(vp_z)^2+ b^2 \nonumber \\
 &+&2\mu b\sqrt{(vp_y)^2+m^2}\Big]^{1/2},
\end{eqnarray}
where $s,\mu=\pm 1$. At $p_x=p_z=0$,
$p_y= \pm \sqrt{b^2-m^2}/v$, formula (\ref{27}) gives the two Weyl points with $\epsilon^W=0$. Note that the bands described by Eq.~(\ref{27}) posses the mirror symmetry with respect to this Weyl-point energy:  $\epsilon_{1,\mu}=-\epsilon_{-1,\mu}$. However, this symmetry is absent for Eqs.~(\ref{5})--(\ref{7}), see Figs.~\ref{fig2}--\ref{fig8}. This asymmetry of the bands is due to the term linear in $\epsilon$ in formula (\ref{6}), and the asymmetry disappears only if $m_2=0$. In this case, equation (\ref{5}) becomes biquadratic with respect to $\epsilon$, and all the energy bands can be found explicitly,
\begin{eqnarray}\label{28}
\epsilon_{s,\mu}&\!=\!&\!s\Big[(t_2p_y)^2+d_z^2+\tilde m^2+m_3^2+m_4^2 \nonumber \\
 &\!+\!&\!2\mu\tilde m\sqrt{\!(t_2p_y)^2\!+\!\frac{(d_zm_1\!-\!m_4m_6\!+\!m_3m_5)^2}{\tilde m^2}}\Big]^{1/2}\!\!\!\!\!\!,
\end{eqnarray}
where $d_z=a'p_{\perp}$ is described by formula (\ref{4}), and $\tilde m\equiv \sqrt{m_1^2+m_5^2+m_6^2}$. Note that at $m_1\neq 0$, Eq.~(\ref{28}) gives the band-contact line, $d_z=a'p_{\perp}= (m_4m_6-m_3m_5)/m_1$, in the plane $p_y=0$. Along this line, one has  $\epsilon_{s,1}=\epsilon_{s,-1}$ for $s=\pm 1$. The appearance of these lines is due to the fail of condition (\ref{16}).

Expressions (\ref{27}) and (\ref{28}) will have similar forms if $d_z$  under the radical in Eq.~(\ref{28}) can be omitted. Therefore, to obtain Eq.~(\ref{27}), we must additionally assume that $m_1=0$. Then, a comparison of Eqs.~(\ref{27}) and (\ref{28}) gives
\[
b=\tilde m=\sqrt{m_5^2+m_6^2},\ \ \  m^2=\frac{(m_4m_6-m_3m_5)^2}{\tilde m^2}.
\]
The term $m_3^2+m_4^2$ in Eq.~(\ref{28}) can be represented as follows:
 \[
 m_3^2+m_4^2=\frac{(m_3m_6+m_4m_5)^2+(m_4m_6-m_3m_5)^2}{m_5^2+m_6^2}.
 \]
Now we take into account that the $m_i$ depend on $p_{\|}$,  and according to Eq.~(\ref{10}), one has $(m_3m_6+m_4m_5)=0$ at $p_{\|}=0$, see Eqs.~(\ref{20}). The parameters $b$ and $m^2$ will be independent of $p_{\|}$ if $\tilde v=v_{\perp}=0$, whereas the last formula in Eqs.~(\ref{20}) leads to
\[
m_3^2+m_4^2=m^2+v_{\|}^2p_{\|}^2.
\]
Eventually, expression (\ref{28}) transforms into a generalization of formula (\ref{27}) in which  $v^2(p_x^2+p_y^2+p_z^2)$  is replaced by $(a'p_{\perp})^2+(t_2p_y)^2+(v_{\|}p_{\|})^2$. Thus, we conclude that the model of Ref.~\cite{jiang} is a special case of a more general spectrum presented in Sec.~\ref{spectrum}, and it is obtained from this spectrum at $m_1=m_2=0$, $a'=t_2=v_{\|}\equiv v$ and at special dependences of the other $m_i$ on the quasimomentum $p_{\|}$ along the nodal line [these dependences correspond to the minimal model described by Eqs.~(\ref{24})].

In Ref.~\cite{moon}, the terms
\[
vt_xp_x+vt_zp_z+vt_yp_y\tau_x
\]
were added to Hamiltonian (\ref{26}). Here the constants $t_i$ satisfy the condition $t_x^2+t_y^2+t_z^2<1$. The first two terms in this expression, in fact, reproduce the last two terms in the formula
\[
\varepsilon=\epsilon+ ap_{\perp}+\frac{d\varepsilon_0}{dp_{\|}}p_{\|}
\]
that follows from the definition $\epsilon=\varepsilon-\bar\varepsilon_0$ and Eq.~(\ref{4}).
As to $vt_yp_y\tau_x$, this term was omitted in Hamiltonian (\ref{1}), (\ref{2}) since as follows from considerations of Ref.~\cite{h.weng}, the velocity $(t_yv)$ has the spin-orbit origin,  and it is relatively small as compared to $t_2$.

\section{Conclusions}\label{conclusions}

We propose the model describing the charge-carriers spectrum near a pair of  close Weyl points in a topological semimetal of the TaAs family. To explain how to choose values for the parameters of this model, we consider NbP and find the values in the cases of the  W1 and W2 points in this material. However, for the W2 points, simplifying condition (\ref{24}) is assumed, and therefore, additional band-structure calculations are required to refine the obtained values. Our analysis  also shows that for the semimetals of the TaAs family, the dispersion of the bands along the certain direction of quasimomentum begins to deviate from the usually assumed linear dependence already in the close vicinity of each of the two Weyl points. This deviation should be taken into account when the Fermi surface and its characteristics are quantitatively analyzed for these materials.

\appendix

\section{Energy spectrum in the immediate vicinity of a Weyl point} \label{A}

In a region of the Brillouin zone near the pair of close Weyl points, let us go from the curvilinear coordinates $p_{\perp}$ and  $p_{\|}$ introduced in Sec.~\ref{spectrum} to the local Cartesian coordinates system $p_{\perp}^{\rm C}$, $p_{\|}^{\rm C}$ that has the origin at the point ($p_{\perp}=0$,$p_{\|}=p_{\|}^W$) of the band-contact line lying in the plane $p_y=0$. Thus, the $p_{\|}^{\rm C}$ axis coincides with the tangent to the band-contact line (\ref{3}) at its point defined by condition (\ref{9}), whereas the coordinate $p_{\perp}^{\rm C}$ is perpendicular to this tangent.
We will also measure the curvilinear coordinate $p_{\|}$ from $p_{\|}^W$. Then, geometrical considerations give the following relation between these two coordinate systems near their common origin:
\begin{eqnarray}\label{a1}
p_{\perp}^{\rm C}&=&(R+p_{\perp})\cos\phi-R, \nonumber \\
p_{\|}^{\rm C}&=&(R+p_{\perp})\sin\phi,
\end{eqnarray}
where the angle $\phi=p_{\|}/R$, and $R$ is the radius of the curvature for the band-contact line at its point $p_{\|}=0$. Assuming $|p_{\|}|, |p_{\perp}|\ll R$, we obtain
\begin{eqnarray*}
p_{\perp}&\approx &p_{\perp}^{\rm C}+\frac{(p_{\|}^{\rm C})^2}{2R}, \\
p_{\|}&\approx &p_{\|}^{\rm C}.
\end{eqnarray*}
We keep the small second term in the expression for $p_{\perp}$ because  it becomes important when $p_{\perp}^{\rm C}\to 0$. Then, instead of formulas (\ref{4}), we arrive at
\begin{eqnarray}\label{a2}
d_z&= &a'p_{\perp}^{\rm C}+\frac{a'(p_{\|}^{\rm C})^2}{2R}, \\
\bar\varepsilon_0&= &\varepsilon_0(0)+\frac{d\varepsilon_0}{dp_{\|}}p_{\|}^{\rm C}+ap_{\perp}^{\rm C}+\frac{1}{2}\frac{d^2\varepsilon_0}{dp_{\|}^2}(p_{\|}^{\rm C})^2+\frac{a(p_{\|}^{\rm C})^2}{2R},~~~ \nonumber
\end{eqnarray}
where we have expanded $\varepsilon_0(p_{\|})$ in powers of $p_{\|}$. The last two terms in the formula for $\bar\varepsilon_0$ can be  neglected for the W2 points, but these terms are important for the W1 points since $(d\varepsilon_0/dp_{\|})=0$ in this case. Below we use the Cartesian coordinates, and for brevity we will omit the index ${\rm C}$ in $p_{\|}^{\rm C}$,  $p_{\|}^{\rm C}$.

Consider the dispersion of the two crossing bands in the immediate vicinity of the Weyl point. In this case,
 \begin{eqnarray}\label{a3}
d_z=-\frac{m_1m_4}{m_6}+\Delta d_z=-\frac{m_1m_4}{m_6}+a'\Delta p_{\perp}+\frac{a'p_{\|}^2}{2R},~~
 \end{eqnarray}
where $\Delta d_z$ and $\Delta p_{\perp}$ are deviations of $d_z$ and $p_{\perp}$ from their values at the Weyl point: $-m_1m_4/m_6$ and $p_{\perp}^{W}$, respectively; $p_{\perp}^W$ is determined by Eq.~(\ref{9}).  Note that the last term in Eq.~(\ref{a3}) may become important if the splitting of the crossing bands is small along the $p_{\|}$ direction. As to the parameters $t_2$, $m_i$ and the combinations $\kappa_i$, they change according to formulas (\ref{20}) when the quasimomentum  shifts relative to the Weyl point.
The term in the brackets in Eq.~(\ref{21}) and all the terms in formula (\ref{22}), are proportional to the squares of the expressions that vanish at the Weyl point.  This means that when the quasimomentum   deviates from the Weyl point,  equation (\ref{21}) looks like the quadratic form relative to  $\Delta\epsilon\equiv \epsilon-(m_2m_4/m_6)$, $\Delta d_z\equiv d_z +(m_1m_4/m_6)$, $\Delta p_y\equiv p_y-p_y^W$, and $p_{\|}$,
\begin{eqnarray}\label{a4}
& &\left[t_2p_y^W(t_2\Delta p_y)-\frac{m_1m_4\Delta d_z}{m_6}-\frac{m_2m_4\Delta\epsilon}{m_6}+c_0p_{\|}\right]^2 \nonumber \\
&=&(m_1^2+m_5^2+m_6^2) \left[\Delta\epsilon+\frac{m_2m_1\Delta d_z}{(m_1^2+m_5^2+m_6^2)}+cp_{\|}\right]^2 \nonumber  \\
&-&\frac{(m_1^2-m_2^2+m_5^2+m_6^2)}{(m_1^2+m_5^2+m_6^2)}(m_5^2+m_6^2)(\Delta d_z+c_2p_{\|})^2  \nonumber \\
&-&\frac{(m_1^2-m_2^2+m_5^2+m_6^2)}{(m_5^2+m_6^2)}(c_1p_{\|})^2,
\end{eqnarray}
where the terms $c_ip_{\|}$ result from the $p_{\|}$ dependences of $t_2$, $m_1$, $m_2$, and $\kappa_i$. These $c_i$ are expressed in terms of $t_2$, $m_i$ corresponding to the Weyl point,  and of the  parameters $v_1$, $v_2$, $\tilde v$, $v_{\perp}$, $v_{\|}$, $t_2'\equiv (dt_2/dp_{\|})_{p_{\|}=0}$ defined in  Eqs.~(\ref{20}),
 \begin{eqnarray}\label{a5}
c_0&=&m_2v_2-m_1v_1+\frac{m_4}{m_6}\sqrt{m_5^2+m_6^2}v_{\perp} \nonumber \\ &-&\frac{1}{2}\sqrt{m_5^2+m_6^2}\Big(1+\frac{m_4^2}{m_6^2}\Big)\tilde v+(p_y^W)^2t_2t_2', \nonumber \\
c_1&=&\sqrt{m_5^2+m_6^2}v_{\|}, \\
c_2&=&\frac{m_4}{m_6}v_1+\frac{m_1}{\sqrt{m_5^2+m_6^2}}(v_{\perp} -\frac{m_4}{m_6}\tilde v), \nonumber \\
c&=&\frac{m_2}{\tilde m^2}\Big(\frac{m_1m_4}{m_6}v_1+ \frac{m_4}{m_6}\sqrt{m_5^2+m_6^2}\tilde v -\sqrt{m_5^2+m_6^2}v_{\perp}\Big) \nonumber \\
&-&\frac{m_4}{m_6}v_2. \nonumber
 \end{eqnarray}
Solving the quadratic equation (\ref{a4}) in $\Delta\epsilon$,
we arrive at
 \begin{eqnarray}\label{a6}
 \Delta\epsilon&=&-\frac{m_1m_2}{q_1}(1-\frac{m_4^2}{m_6^2})\Delta d_z-\frac{m_2m_4t_2^2p_y^W\Delta p_y}{m_6q_1}-v_{0\parallel}p_{\|} \nonumber \\
 &\pm& \Big[\frac{q_2}{q_1^2}(\Delta d_z)^2+2v_{\|,\perp}p_{\|}\Delta d_z+ 2v_{y,\|}p_{\|}(t_2\Delta p_y) \nonumber \\  &-&\!2q_{y,\perp}\Delta d_z(t_2\Delta p_y)\!+\!(v_{\|,\|}p_{\|})^2\! +\!q_{y,y}(t_2\Delta p_y)^2
 \Big]^{1/2}\!\!\!\!,
 \end{eqnarray}
where we have introduced the designations,
 \begin{eqnarray}\label{a7}
 q_1&=& m_1^2+m_5^2+m_6^2-m_2^2\frac{m_4^2}{m_6^2},\\
 q_2&=&\big[m_1^2+m_5^2+m_6^2-m_2^2\big]
 \Big[q_1-m_1^2(1-\frac{m_4^2}{m_6^2})\Big], \nonumber \\
 q_{y,\perp}&=&\frac{m_1m_4t_2p_y^W}{m_6q_1^2}(m_1^2+m_5^2 +m_6^2-m_2^2), \nonumber \\
 q_{y,y}&=&\frac{(m_1^2+m_5^2+m_6^2)(t_2p_y^W)^2}{q_1^2},
\nonumber
 \end{eqnarray}
and the velocities $v_{0\|}$, $v_{\|,\perp}$,  $v_{y,\|}$, and  $v_{\|,\|}$ are  expressible  in terms of $m_i$ and constants $c$, $c_0$, $c_1$, $c_2$ as follows:
 \begin{eqnarray}\label{a8}
v_{0\|}&=&\frac{1}{q_1}(c\tilde m^2+c_0\frac{m_2m_4}{m_6}), \nonumber \\
 v_{y,\|}&=&\frac{t_2p_y^W\tilde m^2}{q_1^2}(c_0+c\frac{m_2m_4}{m_6}), \nonumber \\
 v_{\|,\perp}&=&\frac{\tilde m^2\!\!-\!m_2^2}{q_1}\Big[c_2\!\frac{(m_5^2\!+\!m_6^2)}{\tilde m^2}
 \!-\!\frac{m_1m_4}{m_6q_1}(c_0\!+\!c\frac{m_2m_4}{m_6})\Big], \nonumber \\
 (v_{\|,\|})^2&=&\frac{\tilde m^2-m_2^2}{q_1} \Big[\frac{c_1^2}{m_5^2+m_6^2}+\frac{c_2^2(m_5^2+m_6^2)}{\tilde m^2}\Big] \nonumber \\
 &+&\frac{\tilde m^2}{q_1^2}(c_0+c\frac{m_2m_4}{m_6})^2.
 \end{eqnarray}
It can be shown that the expression under the radical in Eq.~(\ref{a6}) is always nonnegative if conditions (\ref{15}) are fulfilled, and so the  velocities $v_{\|,\perp}$,  $v_{y,\|}$, and  $v_{\|,\|}$  satisfy the following inequalities:
 \begin{eqnarray}\label{a9}
 v_{\|,\|}^2\ge \frac{q_1^2}{q_2}v_{\|,\perp}^2,\ \ \ \
 v_{\|,\|}^2\ge \frac{1}{q_{y,y}}v_{y,\|}^2.
  \end{eqnarray}
Knowing $\Delta\epsilon$, one can find the change in the energy $\Delta\varepsilon\equiv \varepsilon(p_{\perp},p_{\|})-\varepsilon^W$ of any of the crossing bands, using the relationship that follows from the definition of $\epsilon\equiv\varepsilon-\bar\varepsilon_0$,
 \begin{eqnarray}\label{a10}
\Delta\varepsilon=\Delta\epsilon+a\Delta p_{\perp}+ \frac{d\varepsilon_0}{dp_{\|}}p_{\|}+ \frac{1}{2}\frac{d^2\varepsilon_0}{dp_{\|}^2}p_{\|}^2+ \frac{ap_{\|}^2}{2R}.~~~
 \end{eqnarray}
Note that for the direction along $p_{\|}$, formulas (\ref{a3}), (\ref{a6}), (\ref{a10}) contain not only the linear terms in $p_{\|}$ but also the quadratic terms in this component of the quasimomentum. These quadratic terms  can be significant because the velocities $v_1$, $v_2$, $\tilde v$, $v_{\perp}$, $v_{\|}$ associated with the spin-orbit interaction are relatively small in the TaAs family of the semimetals [due to formula (\ref{14}), the term $(p_y^W)^2t_2t_2'$ has the same order of magnitude as the other terms in $c_0$ defined by  the first expression in Eqs.~(\ref{a5})]. However, if a direction, along which the bands are considered, noticeably deviates from  the $p_{\|}$ axis, the quadratic terms in $\Delta d_z$ and $\Delta\varepsilon$ can be omitted. In this situation, it is convenient to rewrite formulas (\ref{a6}), (\ref{a10}) in terms of the variables $\Delta p_x$, $\Delta p_y$ $\Delta p_z$. This representation is  useful when the bands are analyzed along the $p_x$, $p_y$  axes for the W1 points and along the $p_x$, $p_y$, $p_z$ directions in the case of the W2 points.

Let the tangent to the band-contact line lying in the $p_x-p_z$ plane be at the angle $\theta\neq 0$ to the $p_z$ axis near the Weyl point (Fig.~\ref{fig1}).  Then, we have
\begin{eqnarray}\label{a11}
 \Delta\varepsilon&=&\Big[\Big(a-a'\frac{m_1m_2}{q_1}(1- \frac{m_4^2}{m_6^2})\Big)\cos\theta -\varepsilon_0'\sin\theta\Big]\Delta p_x \nonumber \\
 &+&\Big[\Big(a-a'\frac{m_1m_2}{q_1}(1- \frac{m_4^2}{m_6^2})\Big)\sin\theta +\varepsilon_0'\cos\theta\Big]\Delta p_z \nonumber \\
  &-&\frac{m_2m_4t_2^2p_y^W}{m_6q_1}\Delta p_y \pm \sqrt{\sum_{i,j} Q_{ij}\Delta p_i \Delta p_j},
 \end{eqnarray}
where $\varepsilon_0'\equiv (d\varepsilon_0/dp_{\|})-v_{0\|}$ is the renormalized value of $d\varepsilon_0/dp_{\|}$; $i,j=x,y,z$, and we take $\Delta d_z=a'\Delta p_{\perp}$, neglecting the term  $a'p_{\|}^2/(2R)$ here. The symmetric matrix $Q_{ij}$ has the following elements:
 \begin{eqnarray}\label{a12}
Q_{xx}&=&\frac{q_2}{q_1^2}(a')^2\cos^2\!\theta+v_{\|,\|}^2\sin^2\!\theta -2a'v_{\|,\perp}\sin\theta \cos\theta, \nonumber \\
Q_{zz}&=&\frac{q_2}{q_1^2}(a')^2\sin^2\!\theta+v_{\|,\|}^2\cos^2\!\theta +2a'v_{\|,\perp}\sin\theta \cos\theta, \nonumber \\
Q_{yy}&=&\frac{(m_1^2+m_5^2+m_6^2)(t_2p_y^W)^2t_2^2}{q_1^2},\\
Q_{xz}&=&Q_{zx}=\Big(\frac{q_2}{q_1^2}(a')^2-v_{\|,\|}^2\Big)\frac{\sin(2\theta)}{2} +a'v_{\|,\perp}\cos(2\theta), \nonumber \\
Q_{xy}&=&Q_{yx}=-t_2(v_{y,\|}\sin\theta+a'q_{y,\perp}\cos\theta ), \nonumber \\ Q_{yz}&=&Q_{zy}=t_2(v_{y,\|}\cos\theta-a'q_{y,\perp}\sin\theta ). \nonumber
 \end{eqnarray}
If the matrix $Q_{ij}$ and the coefficients before $\Delta p_i$ ($i=x,y,z$) in formula (\ref{a11}) are known from the the band-structure calculations, one can find the parameters $a$, $\varepsilon_0'$, $a'$, $v_{\|,\|}$, $v_{\|,\perp}$, $v_{y,\|}$ and $m_1$, using the coefficients before $\Delta p_x$, $\Delta p_z$, and $Q_{xx}$, $Q_{zz}$, $Q_{xz}$, $Q_{yx}$, $Q_{yz}$. (The coefficients before $\Delta p_y$ and $Q_{yy}$ have already been used in determining $m_2$, $\tilde m$, $m_4/m_6$, $t_2$ from the dispersion of the bands along the $p_y$ axis, Sec.~\ref{NbP}.) However, the band-structure calculations usually give only the coefficients before $\Delta p_i$ and $Q_{xx}$, $Q_{zz}$, $Q_{yy}$. For example, the $v_{x\pm}$ and $v_{z\pm}$ presented in Table \ref{tab3} yield
 \begin{eqnarray*}
  Q_{xx}&=&\Big(\frac{v_{x+}-v_{x-}}{2}\Big)^2=(1.85\times10^5 {\rm m/s})^2, \\ Q_{zz}&=&\Big(\frac{v_{z+}-v_{z-}}{2}\Big)^2=(2.4\times10^5 {\rm m/s})^2,
\end{eqnarray*}
whereas the coefficients before $\Delta p_x$ and $\Delta p_z$ are $0.5(v_{x+}+v_{x-})=0.25\times 10^5$ m/s and $0.5(v_{z+}+v_{z-})=1.4\times 10^5$ m/s, respectively.

In the case of the W1 points, one has $\theta=0$, the $p_{\perp}$ coincides with the $p_x$ axis and the $p_{\|}$ is along the $p_z$ direction (Fig.~\ref{fig1}). Then, the terms proportional to $p_{\|}^2$ in  $\Delta d_z$ and in Eq.~(\ref{a10}) can manifest themselves in the charge-carriers dispersion along the $p_z$ axis (Sec.~\ref{W1}).

\section{Minimal model for $m_i(p_{\|})$} \label{B}

Let us denote the left-hand side of Eq.~(\ref{21}) as the function $F(\epsilon,p_y^2t_2^2,d_z,m_1,m_2,\tilde\kappa,\kappa_{\perp},\kappa_{\|})$.
For the band $\varepsilon_i(p_y,p_{\perp},p_{\|})$, the change $\Delta\epsilon_i\equiv \epsilon_i(p_{\|})-\epsilon_i(0)$ of the quantity $\epsilon_i\equiv \varepsilon_i-\bar\varepsilon_0$
at fixed  $p_y=p_{\perp}=0$ or at $p_y=p_y^W$, $p_{\perp}= p_{\perp}^W$ can be found from the following formula \cite{com3}:
 \begin{eqnarray}\label{b1}
  \frac{\partial F}{\partial \epsilon}\Delta\epsilon_i\!&=&\! -2\frac{\partial F}{\partial (p_y^2t_2^2)}p_y^2t_2t_2'p_{\|}\!-\! \frac{\partial F}{\partial d_z}\frac{a'p_{\|}^2}{2R}\!-\! \sum_{j=1}^{2}\frac{\partial F}{\partial m_j}v_jp_{\|}~~
  \nonumber \\
  &-&\!\!\sum_{l}\frac{\partial F}{\partial \kappa_l}v_lp_{\|},~~~~
   \end{eqnarray}
where $\kappa_l=\tilde\kappa$, $\kappa_{\perp}$, $\kappa_{\|}$ (and $v_l=\tilde v$, $v_{\perp}$, $v_{\|}$), the partial derivatives of the function $F$  are calculated for given values of $p_y$, $p_{\perp}$ and corresponding $\epsilon_i= \epsilon_i(p_y,p_{\perp},p_{\|}=0)$, $d_z=d_z(p_{\perp},p_{\|}=0)$, $m_j=m_j(p_{\|}=0)$, $\kappa_l=\kappa_l(p_{\|}=0)$, $t_2=t_2(p_{\|}=0)$. This formula is expected to be valid over an interval of $p_{\|}$ comparable with the distance between the close Weyl points since the factors near the partial derivatives on its right-hand side are relatively small in this interval. The $\Delta \epsilon_i$ thus calculated together with formula (\ref{a10}) at $\Delta p_{\perp}=0$ give $\Delta\varepsilon_i$. If these $\Delta\varepsilon_i$ are also found from the band-structure calculations, a set of equations (\ref{b1}) for different $i$ makes it possible to find values of the parameters $v_1$, $v_2$, $\tilde v$, $v_{\perp}$, $v_{\|}$, $t_2'$ of the model spectrum.

Small values of  $v_1$, $v_2$, $\tilde v$, $v_{\perp}$, $v_{\|}$, and $p_y^Wt_2'$ lead to small velocities $v_{0\|}$, $v_{\|,\perp}$,  $v_{y,\|}$,  $v_{\|,\|}$ defined in Appendix \ref{A}. In this case, the $v_{0\|}$ produces a little renormalization of $d\varepsilon_0/dp_{\|}$, whereas the  $v_{\|,\perp}$,  $v_{y,\|}$ induce a little deviation of the direction, along which the splitting of the crossing bands is minimal, from the $p_{\|}$ axis, i.e., from the direction of the tangent to the band-contact line. These small quantitative effects do not result in a noticeable change in the dispersion of the bands. In the first approximation, one can neglect the velocities $v_{0\|}$, $v_{\|,\perp}$,  $v_{y,\|}$ and take into account only $v_{\|,\|}$, which determines the qualitative feature of the spectrum, the splitting of the bands along the $p_{\|}$ direction. In this approximation, one obtains $c=c_0=c_2=0$ from expressions (\ref{a8}) for $c_i$, and we may use conditions (\ref{24})  in formulas (\ref{20}). Equations (\ref{a8}) also give   $c_1=v_{\|,\|}\sqrt{(m_6^2+m_6^2)q_1/(\tilde m^2-m_2^2)}$, and with Eqs.~(\ref{a5}), we arrive at
 \begin{eqnarray}\label{b2}
v_{\|}=v_{\|,\|}\sqrt{\frac{q_1}{(\tilde m^2-m_2^2)}}\,.
 \end{eqnarray}
The use of Eqs.~(\ref{20})--(\ref{23}) with conditions (\ref{24}) and formula (\ref{b2}) lead to the  minimal model of the four-band spectrum (Sec.~\ref{spectrum}) that reproduces the main feature of the Weyl points,  the linear splitting of the bands in all the  directions of the quasimomentum ${\bf p}$ at ${\bf p}\to {\bf p}^W$.

Within this minimal model,  Eqs.~(\ref{a11}) and (\ref{a12}) lead to the simple relationships determining the parameters $a$, $a'$, $d\varepsilon_0'/dp_{\|}$ and $v_{\|,\|}$ in terms of $v_{x\pm}$ and $v_{z\pm}$ presented in Table \ref{tab3},
 \begin{eqnarray}\label{b3}
v_{\|,\|}^2&=&\frac{Q_{zz}-Q_{xx}\tan^2\theta}{1-\tan^2\theta}, \nonumber \\
(a')^2&=&\frac{q_1^2}{q_2}\frac{(Q_{xx}-Q_{zz}\tan^2\theta)}{(1- \tan^2\theta)},  \\
a&=&a'\frac{m_1m_2}{q_1}(1-\frac{m_4^2}{m_6^2}) \nonumber \\
&+&\frac{v_{z+}^{W2}+v_{z-}^{W2}}{2}\sin\theta+\frac{v_{x+}^{W2}+ v_{x-}^{W2}}{2}\cos\theta, \nonumber \\
\frac{d\varepsilon_0}{dp_{\|}}\!&=&\!\frac{v_{z+}^{W2}+ v_{z-}^{W2}}{2}\cos\theta\!- \frac{v_{x+}^{W2}+ v_{x-}^{W2}}{2}
\sin\theta, \nonumber
 \end{eqnarray}
where $Q_{xx}=(v_{x+}^{W2}-v_{x-}^{W2})^2/4$, $Q_{zz}=(v_{z+}^{W2}-v_{z-}^{W2})^2/4$, and $q_1$, $q_2$ are defined by Eqs.~(\ref{a7}). The formula for $v_{\|,\|}^2$ shows that in the minimal model, there is a critical value $\theta=\theta_0$,
 \begin{eqnarray}\label{b4}
 \tan^2(\theta_0)=\frac{Q_{zz}}{Q_{xx}},
 \end{eqnarray}
at which the velocity $v_{\|,\|}$ becomes zero. However,  at $|\theta|<|\theta_0|$, $v_{\|,\|}$ sharply increases, $v_{\|,\|}\propto \sqrt{|\theta_0|-|\theta|}$.
For example, for $v_{x\pm}$, $v_{z\pm}$ from Table \ref{tab3}, we find $|\theta_0|\approx 128^{\circ}$, whereas at $\theta=125^{\circ}$, $v_{\|,\|}$ already reaches $ 1.08\times 10^5$ m/s. Thus, a closeness of $\theta$ to $\theta_0$ is the necessary condition for the applicability of the minimal model to the description of the four-bands spectrum.

\section{Cross-sectional areas of Fermi surface near Weyl points}\label{C}

\begin{figure}[t] % %%%%%%%%%%%%%%%%%%%%%%%%%%%%%%%%%%%%%
 \centering  \vspace{+9 pt}
\includegraphics[scale=0.99]{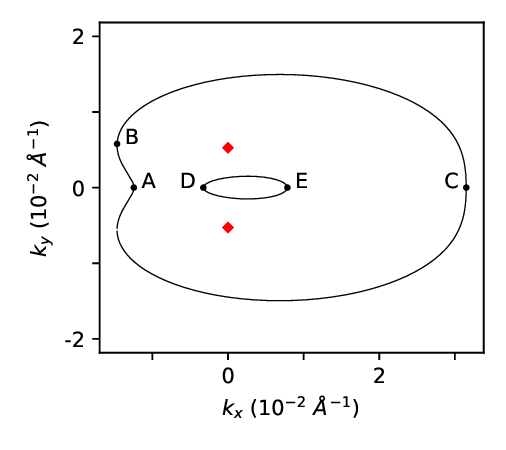}
\caption{\label{fig9} Cross sections of the Fermi surface  for the third (outer curve) and  forth (inner curve) bands of NbP by the plane perpendicular to the $p_z$ axis and passing through the W1 points (red rhombi). Here $k_x=p_x/\hbar$, $k_y=p_y/\hbar$, which are measured from the middle of the line segment connecting the Weyl points. The Fermi energy is shifted by $24$ meV downwards compared to its value in Table \ref{tab2} and Fig.~\ref{fig2} in order to explain the notations $k_y^{(1a)}(k_x')$, $k_y^{(1b)}(k_x')$, $k_y^{(2)}(k_x')$ introduced in the text [$k_x'$ coincides with $k_x$ for the plane perpendicular to $k_z$]. The upper segment D-E of the closed curve shows nonzero values of $k_y^{(1b)}(k_x)$, whereas the segments B-A and B-C  depict the functions $k_y^{(1a)}(k_x)$ and $k_y^{(2)}(k_x)$, respectively, in the regions where their imaginary parts vanish.
}
\end{figure}

Equations (\ref{20})--(\ref{23}) permit one to calculate the cross-sectional areas $S$ of the Fermi surface by the planes which are perpendicular to the magnetic field ${\bf H}=H{\bf n}$ and located in the vicinity of the Weyl points. Here ${\bf n}$ is the unit vector specifying the direction of the magnetic field, and for simplicity, we consider only the case  when ${\bf n}$ lies in the $p_x$-$p_z$ plane, i.e., ${\bf n}=(n_x,0,n_z)$. Then, the plane perpendicular to ${\bf H}$ is defined by the equation,
 \begin{eqnarray}\label{c1}
 p_xn_x+p_zn_z=p_0,
 \end{eqnarray}
where the quasimomentum ${\bf p}=(p_x,0,p_z)$ is reckoned from the origin of the Cartesian coordinate system described in Appendix \ref{A}, and the constant $p_0$ is the distance between this origin and the  plane. The Cartesian coordinates $p_{\perp}$ and $p_{\|}$, which are used in formulas (\ref{20})--(\ref{23}), are expressed in terms of the $p_x$ and $p_z$ as follows:
 \begin{eqnarray}\label{c2}
 p_{\perp}&=&\cos\theta p_x+\sin\theta p_z,
   \nonumber \\
 p_{\|}&=&-\sin\theta p_x +\cos\theta p_z,
  \end{eqnarray}
where $\theta$ is the angle between the tangent to the band-contact line at the point $p_x=0$, $p_z=0$ and the $p_z$ axis, Fig.~\ref{fig1}.
To calculate the area $S$, it is convenient to introduce also the coordinates $p_x'$ and $p_z'$ where $p_z'$ is along the direction of the magnetic field $H$, whereas $p_x'$ is parallel to the plane of the cross section,
 \begin{eqnarray}\label{c3}
p_z'&=& n_xp_x + n_zp_z,  \nonumber \\
p_x' &= &n_zp_x - n_xp_z.
 \end{eqnarray}
Using Eqs.~(\ref{c2}) and (\ref{c3}), we can express the $p_{\perp}$ and $p_{\|}$ via the $p_x'$ and $p_z'=p_0$,
 \begin{eqnarray}\label{c4}
p_{\perp}&=& (\cos\theta n_x + \sin\theta n_z) p_0 + (\cos\theta n_z  - \sin\theta n_x ) p_x',~~\\
p_{\|} &=& (-\sin\theta n_x + \cos\theta n_z) p_0 - (\sin\theta n_z + \cos\theta n_x) p_x'.~~\nonumber
 \end{eqnarray}
Then, Eqs.~(\ref{20})--(\ref{23}) with $\epsilon=\varepsilon_F- \bar\varepsilon_0$ give the explicit dependence of $p_y$ on $p_x'$. This dependence describes the intersection of the Fermi surface and plane (\ref{c1}), Fig.~\ref{fig9}.

   %%%%%%%%%%%%%%%%%%%%%%%%%%%%%%%%%%%%%%%%%%

To calculate the area $S$ of this cross section, it is sufficient to consider only the two branches $p_y^{(i)}(p_x')$ ($i=1,2$) of the complex-valued function $p_y(p_x')$, the real parts of which are  positive, Re$[p_y^{(i)}(p_x')]>0$. Here the $p_y^{(1)}(p_x')$ and $p_y^{(2)}(p_x')$ correspond to $-\sqrt{Y}$ and $+\sqrt{Y}$ in Eq.~(\ref{21}), respectively. In the general case,  the branch $p_y^{(1)}(p_x')$ has several segments of the $p_x'$ axis, inside which Im$[p_y^{(1)}(p_x)]=0$, Fig.~\ref{fig9}. If, at least at one of the ends of such a segment, the branch $p_y^{(1)}(p_x')$ merges with $p_y^{(2)}(p_x')$, this portion of the $p_x'$ axis will be designated as the segment of the type $a$. However, the segment is denoted as the $b$ segment if $p_y^{(1)}=0$ at both its ends. We now define the following function $p_y^{(1a)}(p_x')$: $p_y^{(1a)}(p_x') =p_y^{(1)}(p_x')$ for all $p_x'$ outside the $b$ segments, and $p_y^{(1a)}(p_x')\equiv 0$ inside them. Then, the real-valued function  $p_y^{(1b)}(p_x')\equiv p_y^{(1)}(p_x') -p_y^{(1a)}(p_x')$ differs from zero only inside the $b$ segments. Eventually, we obtain the following expression for the area $S$ of the orbit in the third band:
  \begin{eqnarray}\label{c5}
  S=\!\frac{2\pi^2\hbar^2 F}{\phi_0}\!=2\!\int\!\!\Big({\rm Re}[p_y^{(2)}\!(p_x')]\!-\!{\rm Re}[p_y^{(1a)}\!(p_x')]\Big)dp_x',~~~~~
  \end{eqnarray}
where $\phi_0=2.07\times 10^{-15}$ Tm$^2$ is the flux quantum, $F$ is the frequency of the quantum oscillations associated with this cross-sectional area, and the integration is carried out over any interval of $p_x'$ that covers the width of the cross section in this direction. [For points outside this interval, ${\rm Re}(p_y^{(1a)})={\rm Re}(p_y^{(2)})$. Therefore, these points do not contribute  to the integral, and the size of the interval is unimportant for the calculation.] The cross-sectional area of the orbit in the fourth band is given by
  \begin{eqnarray}\label{c6}
  S=\!\frac{2\pi^2\hbar^2 F}{\phi_0}=2\!\int\!\!p_y^{(1b)}(p_x')dp_x'.~~~~~
  \end{eqnarray}
Note that formulas (\ref{c5}), (\ref{c6}) are also applicable for calculating the cross-sectional areas in the case of the hole Fermi surface. In this case, formula (\ref{c5}) gives the area of the hole orbit in the second band, whereas Eq.~(\ref{c6}) corresponds to the hole orbit in the first band.

In NbP, the difference  $\varepsilon_F-\varepsilon^{W1}$ exceeds the energy barrier separating the W1 points \cite{lee,klotz}, Fig.~\ref{fig2}. In this case, in the magnetic field $H\parallel p_z$, the electron orbit of the third band surrounds the two close W1 points. The area of this orbit, Eq.~(\ref{c5}),  determines the so-called frequency $F_{\alpha1}$ of the quantum oscillations, $F_{\alpha1}=S_{\alpha1}\phi_0/(2\pi^2\hbar^2)$ \cite{klotz}.
Formula (\ref{c6}) gives the so-called frequency $F_{\alpha 2}$ \cite{klotz} associated with the forth band.


\begin{thebibliography}{}
\bibitem{com} An analysis shows that the condition $m_1^2+m_5^2+m_6^2=m_2^2$, which vanishes $Y_{min}$, does not lead to the appearance of the Weyl points.
\bibitem{com1} Formulas (\ref{5})-(\ref{7}) are invariant under the following two transformations of $m_i$: (i) $m_4\to -m_4$ and $m_6\to -m_6$, (ii) $m_6\to -m_6$, $m_3\to -m_3$, $m_2\to -m_2$, and $m_1\to -m_1$. Therefore, without the loss in generality, we may set $m_6>0$, $m_4>0$.
\bibitem{com2} Equations (\ref{5})--(\ref{7}) remain invariant if the signs of $m_1$ and $d_z$ (i.e., $a'$) change simultaneously, and so without the loss in generality, we assume  below that $a'$ is always positive, whereas $m_1$ can have either sign.
\bibitem{comm} For $\kappa_{\|}$, one has $\kappa_{\|}(p_{\|})-\sqrt{m_5^2+m_6^2}v_{\|}p_{\|}\propto p_{\|}^3$. In the other formulas of set (\ref{20}), the additional terms are proportional to $p_{\|}^2$, whereas the linear terms vanish.
\bibitem{com3} If $p_y=p_y^W$, $p_{\perp}=p_{\perp}^W$, formula  (\ref{b1}) fails for the crossing bands ($i=2$, $3$) since one has $(\partial F/\partial \epsilon)=0$ for these bands. In this case, Eq.~(\ref{a6}) has to be used for the calculation of $\Delta \epsilon_i$ with $i=2$, $3$.
\end{thebibliography}
\end{document}